\documentclass[11pt]{article} 

\usepackage{graphicx}
\usepackage{amsfonts,amsmath,graphicx,color,epsfig}
\usepackage{todonotes}
\usepackage{changes}
\usepackage{epsfig}
\usepackage{amsmath}
\usepackage{amsthm}
\usepackage{amssymb}
\usepackage{latexsym}
\usepackage{color}
\usepackage{mathrsfs}
\usepackage{natbib}
\usepackage{hyperref}
\usepackage[toc,page]{appendix}
\usepackage[left=2.5cm,top=4cm,right=2.5cm,bottom=3cm]{geometry}
\usepackage{float}
\usepackage[font=small,labelfont=bf]{caption}
\usepackage{subcaption}
\usepackage{subfloat}
\newcommand{\dW}{{\partial W}}

\newcommand{\dd}{\, \text{d}}
\newcommand{\de}{\, \text{d}}

\newcommand{\bP}{{\mathbb{P}}}

\newcommand{\bR}{{\mathbb{R}}}
\newcommand{\bE}{{\mathbb{E}}}

\newcommand{\lck}{\left\lbrack}
\newcommand{\rck}{\right\rbrack}
\newcommand{\lce}{\left\lbrace}
\newcommand{\rce}{\right\rbrace}

\newcommand{\var}{\mathbb{V}\mathrm{ar}}
\newcommand{\cov}{\mathbb{C}\mathrm{ov}}

\newcommand{\un}{\mathbb{I}}

\newcommand{\resub}[1]{{\color{black}#1}}

\begin{document}
\begin{center}
{\Large Mapping the intensity function of a non-stationary point process

in unobserved areas}

\bigskip

Edith \textsc{Gabriel}\footnote{\tt{edith.gabriel@inrae.fr}}$^a$, Francisco \textsc{Rodr\'{i}guez-Cort\'{e}s}$^b$, J{\'e}r{\^o}me \textsc{Coville}$^a$,

Jorge \textsc{Mateu}$^c$ and Jo{\"e}l \textsc{Chad{\oe}uf$^d$}

 \bigskip

{\small
$^a$ Biostatistics and Spatial Processes Unit, INRAE, F-84911 Avignon, France

$^b$ Escuela de Estad\'istica, Universidad Nacional de Colombia, Medell\'in, Colombia

$^c$ Department of Mathematics, University Jaume I, Castell\'{o}n, Spain

$^d$ Statistics, UR1052, INRA, F-84911 Avignon, France
}

\bigskip


\begin{minipage}{12.5cm}
  {\small
\textsc{Abstract}
Seismic networks provide data that are used as basis both for public safety decisions and for scientific research. Their configuration  affects the data completeness, which in turn, critically affects several seismological scientific targets (e.g., earthquake prediction, seismic hazard...).
In this context, a key aspect is how to map earthquakes density in seismogenic areas from censored data or even in areas that are not covered by the network. We propose to predict the spatial distribution of earthquakes from the knowledge of presence locations and geological relationships, taking into account any \resub{interaction} between records.
Namely, in a more general setting, we aim to estimate the intensity function of a point process, conditional to its censored realization, as in geostatistics for continuous processes. We define a predictor as the best linear unbiased combination of the observed point pattern. We show that the weight function associated to the predictor is the solution of a Fredholm equation of second kind. Both the kernel and the source term of the Fredholm equation are related to the first- and second-order characteristics of the point process through the intensity and the pair correlation function. Results are presented and illustrated on simulated non-stationary point processes and real data for mapping Greek Hellenic seismicity in a region with unreliable and incomplete records.

\medskip

\textsc{keywords}: Conditional intensity, Earthquakes, Fredholm equation, Non-stationarity, Second-order characteristics, Spatial point processes.

}
\end{minipage}

\end{center}

\section{Introduction}
\label{sec:introduction}


Mapping is a key issue in environmental science. A common and first example lies in ecology when mapping species distribution.
When the location of individuals is known, we estimate the local density (usually by kernel smoothing), the so-called intensity in point process theory. However, point locations are usually unreachable at the survey scale, so that sampling methods are used; distance sampling or quadrat sampling approaches are possibly the most common ones. When no covariate is available, a \resub{mean} density estimation is then performed.

However, species distribution characteristics vary spatially as they are governed by environmental data.
Several approaches have been developed in that way for species data formed by reported presence locations, also called occurrence-only records (pure records of locations where a species occurred).
Generally called Species Distribution Models (SDM), they aim to explain species occurrences from environmental variables.
If they are used to gain ecological and evolutionary insights, they are also widely used for model-based interpolation across landscapes or to predict distributions to new geographic regions~\citep{elith2009}. SDMs are often based on multivariate statistical analysis methods, such as Generalized Linear/Additive Models (GLM/GAM).
The most popular models are Maxent~\citep{philipps2006} and Maxlike~\citep{royle2012}. In these models point locations are aggregated to grid cells and whether one or several individuals are observed in a cell, a one is recorded. Then, their aim is to estimate occurrence probability (species' probability of presence in a grid cell) maps. See e.g.~\cite{merow2014} for a comparison of the two models and recommendations about their use.


 Point process models offer a natural framework for species distribution modelling.
Key concerns about SDMs lie in the loss of significant information about the spatial distribution when aggregating point locations, and the dependence of the results on the spatial resolution~\citep{renner2013}.
Although point process models are connected to Maxent~\citep{aarts2012,renner2013}, they use a continuous landscape rather than a discretized one, and the number of records is observed and comes from a random process rather than fixed (number of cells/quadrats). ~\cite{renner2015} showed that using point process models presents many advantages, including some clarification about the response variable and model assumptions which, in addition, can be checked. Furthermore, because they operate at the individual level, point process models can incorporate interaction between individuals and dependence to environmental covariates.

Another concern about all these approaches is that they are Poisson model-based: their intrinsic definition does not account for relationship between individuals.
These interactions nevertheless exist. Competition among individuals often leads to empty areas around each point, so-called exclusion by distance, mimicked by inhibition models. The American Redstarts, for example, compete with conspecifics for habitat in their winter grounds: the re-occupation of the empty areas supports the hypothesis that territoriality in this species acts to exclude conspecifics from certain winter habitat~\citep{MSH1993}.
On the contrary, individuals can be arranged by groups as with gregarious animals, such groups can also describe some local dispersion of the species around parents (as with plants). This arrangement is achieved in cluster models. The Shorea congestiflora is a dominant species in a 25-‐ha forest dynamics plot in a rain forest at Sinharaja (Sri Lanka), which apparently shows clustering at several scales~\citep{WTGSGNOT07}. These effects can be mixed, with individuals arranged in groups, but at certain distance of each other inside each group. This can be the case for the spatial distribution of Northern Gannet~\citep{chadoeuf2011}.

\medskip

Similar issues arise in environmental science.
The case we consider in this paper concerns seismic mapping in the Aegean Sea region. This is an area with high seismic density due to the presence of numerous volcanoes and plate movements. Earthquakes have a heterogeneous spatial distribution and we might be interested in understanding why certain regions are more favorable than others. Modeling earthquakes will therefore require to take into account geological information and interactions between events, clustering being often linked to aftershocks. Evaluating the seismic hazard requires a reliable monitoring network and sufficient coverage, what is rarely the case, and a lack of recording  may not be due to the absence of an earthquake, but to an insufficient or unreliable network. The question then arises of mapping seismic activity in areas where the observation is unreliable or even when the area is not covered by the network, in order to access a relevant map where the seismicity is particularly high.

\medskip 

From a statistical point of view, the method developed in~\cite{gabriel2017} aimed to predict the local \resub{variations in the} intensity of a spatial point process accounting for the individual relationships modeled by the pair correlation function
(which is related to the probability to find a second point of the process at a given distance from a known point of the process). The main interest of this approach is that it estimates, using only first- and second-order characteristics of the point process, the local intensity outside the observation window, hereafter called prediction. The prediction of the local intensity is obtained conditionally to the records in the observation window.
However, this method did not allow to consider environmental covariates and thus
did not take into account potential spatial variations driven by these covariates at large scale, which may lead to unrealistic predictions in ecology and environment.

\medskip

To fill this potential unwished situation, we propose to predict the spatial distribution of earthquakes from the knowledge of presence locations and geological relationships, taking into account any \resub{interaction} between records.
Namely, in a more general setting, we aim to estimate the intensity function of a point process, conditional to its censored realization, as in geostatistics for continuous processes. We define a predictor as the best linear unbiased combination of the observed point pattern, where the weight function associated to the predictor is the solution of a Fredholm equation of second kind related to the first- and second-order moments of the point process.
We describe our approach in~Section~\ref{sec:prediction}.
We evaluate the goodness of our predictions through a simulation study (Section~\ref{sec:simulation}) for several cluster models. In Section~\ref{sec:application}, our methodology is applied to predict and map Hellenic seismicity in a region with unreliable and incomplete records.

\section{Predicting the intensity conditionally to the observation}
\label{sec:prediction}

We consider in the following a spatial point process $\Phi$ in $\bR^2$, i.e. a random pattern of points for which both the number of  points  and  their  locations are random. Let us denote $\Phi (B)$ the number of points of $\Phi$ in any \resub{Borelian} set $B$, and $\Phi_B$ their locations in $B$.
\resub{We assume that the point process $\Phi$ is simple (i.e. the probability of all points of $\Phi$ being distinct is one) and that it has a density probability with respect to the unit rate Poisson process $Y$.}

The intensity function, denoted by $\lambda (x)$, is defined as the function such that
$\int_B \lambda (x) \dd x = \bE \lck \Phi (B) \rck$, for $B$ any \resub{Borelian} set; this corresponds to the local probability to observe a point of $\Phi$ at a fixed location
(if $\dd x$ is an infinitesimal volume around location $x$, then \resub{$\bP \lck \Phi(\dd x)=1 \rck /  \dd x = \lambda (x)$ as $\dd x \to 0$}).
The intensity function provides a trend in the spatial variation of points density, and we suppose that the intensity
is driven by spatial covariates $Z$, such that $\lambda(x)=h(Z_x)$, where both $h$ and $Z$ are known.
The interaction between points is described through the pair correlation function $g(x,y)$, which gives the \resub{extent to} which the probability to find a point  at a location $y$ changes by the presence of a point of the process at location $x$ (if $\dd y$ is an infinitesimal volume around $y$, then \resub{$\bP \lck \Phi (\dd y)=1 \mid x \in \Phi \rck /\dd y = g(x,y)\lambda (y)$ as $\dd y \to 0$}).
We assume that the process is second-order intensity reweighted stationary~\citep{baddeley2000}. This assumption means
that its intensity varies in space, but the pair correlation function between two locations depends only on their difference vector.

Here we consider $W \subset \bR^2$, a window of interest, and we assume that $\Phi$ has only been observed in some observation window $W_{obs} \subset W$.
\resub{To predict the remaining point process $\Phi \cap \{W \backslash W_{obs}\}$, \cite{coeurjolly2017}  express
the conditional distribution of $\Phi \cap \{W \backslash W_{obs}\}$ given $\Phi\cap W_{obs}= \Phi_{W_{obs}}$  in terms of the conditional density
$$f_{W \backslash W_{obs}} \left(\Phi_{W \backslash W_{obs}} | \Phi_{W_{obs}} \right) = f \left( \Phi_{W_{obs}} \cup \Phi_{W \backslash W_{obs}} \right) / f_{W_{obs}} \left( \Phi_{W_{obs}} \right), $$
w.r.t $Y_{W \backslash W_{obs}}$ and where $f_{W_{obs}}$ is the marginal density of $\Phi \cap {W_{obs}}$ w.r.t $Y_{W_{obs}}$.} Unfortunately, the density of $\Phi$ restricted to $W_{obs}$ is barely tractable (and hard to handle) except for just some few processes, such as Poisson, Gibbs and determinantal processes;  however, this is not the case for Cox and cluster point processes that are often used to model environmental or ecological point patterns.

Our aim is \resub{not to predict $\Phi \cap \{W \backslash W_{obs}\}$ but rather to estimate its local intensity variations at any location $x_o \in W \backslash W_{obs}$ conditionally to $h$, $Z$ and $\Phi_{W_{obs}}$.
 Following~\cite{gabriel2016,gabriel2017}, we refer to it as the spatial \emph{local intensity} of $\Phi$ , that we define by the following limit
$$\lambda(x_o|\Phi\cap W_{obs} = \Phi_{W_{obs}})=\lim\limits_{\nu(\dd x_o) \to 0}\frac{\bE \lck \Phi \cap \dd x_o|\Phi\cap W_{obs}= \Phi_{W_{obs}}\rck}{\nu(\dd x_o)}.$$
Hereafter, we denote the conditioning $\Phi \cap W_{obs} = \Phi_{W_{obs}}$, by ``$| \Phi_{W_{obs}}$''.
 We predict the local intensity at points $x_o \in W \backslash W_{obs}$ using a linear predictor mimicking kriging, it can be written as
\begin{equation}\label{weightseq}
\widehat{\lambda}(x_o|\Phi_{W_{obs}})=
\int_{W_{obs}} w(x;x_o) \sum_{y \in \Phi_{W_{obs}}} \delta(x-y) \dd x =
\sum_{x \in \Phi_{W_{obs}}} w(x;x_o),
\end{equation}
where $\delta$ denotes the Dirac delta function and $w$ is a weight function 
solution of the Fredholm equation of the second kind (see Apprendix~\ref{app:fred} for details)~:
\begin{align}\label{Fredhomeq}
w(x;x_o) + \displaystyle\int_{W_{obs}} w(y;x_o)  k(x,y) \dd y =  f(x;x_o),
\end{align}
with kernel
$$k(x,y) = \lambda(y) \left( g(x-y) - \dfrac{1}{\nu(W_{obs})}
\displaystyle \int_{W_{obs}} \lambda(x) g(x-y) \dd x \right)$$
and source term
$$f(x;x_o) = \lambda(x_o) \left(\dfrac{1}{\nu(W_{obs})} + g(x-x_o) - \dfrac{1}{\nu(W_{obs})}
\displaystyle \int_{W_{obs}} \lambda(x) g(x-x_o) \dd x  \right).$$
The solution of this Fredholm equation is implicit but it can be numerically approximated using the  Galerkin approximation method \citep{kress2013}.
Briefly, let ${\mathcal T}_h$ be a given mesh partitioning $W_{obs}$, e.g obtained by triangulation, and consider $\{\varphi_i\}_{i=1,\dots,N}$ a basis of the approximation space $V_h$,  $N=\dim V_h$. We approximate the weight function as a combination of the basis elements~: $w(x)\approx \sum_{i=1}^N w_i\varphi_i(x)$.
This approximation plugged into the Fredholm equation leads to a linear problem for all $\varphi_i$:
$$
  \sum_{j=1}^N w_j\int_{W_{obs}}\left(\varphi_i(x)\varphi_j(x)+\int_{W_{obs}}\int_{W_{obs}}k(x,y)\varphi_j(y)
\varphi_i(x) \dd y \right)=\int_{W_{obs}}f(x;x_o)\varphi_i(x) \dd x.
$$
 Introducing a matrix formulation of the previous equation reformulates as the so-called Galerkin equation,
 $Mw + Kw = F$
  where $M$ is the mass matrix, $K = \left( \int \int k(x,y)\varphi_i(x)\varphi_j(y) \dd x \dd y \right)_{i,j}$ and
$F=\left( \int f(x;x_o)\varphi_i(x) \dd x\right)_{i}$.
That can be simplified by
\begin{equation}\label{eq:galerkin}
(Id+\mathcal{K} M)w=M^{-1}F
\end{equation}
using some projection on the approximation space $\mathcal{K}=(\mathcal{K}_{lm})_{l,m}$ and
$K=M\mathcal{K}M$. We finally get the weights from Equation~(\ref{eq:galerkin}).
In all the illustrations that follow we use this approach to approximate the weight functions and make the prediction.}

\section{Simulation study}
\label{sec:simulation}
The objective of this section is twofold. First, we want to visualize how the prediction is affected by both the distance from the prediction point $x_o$ to the observed window $W_{obs}$ and the point process structure, and we want to measure the variability between our estimator $\widehat{\lambda} (x_o | \Phi_{W_{obs}})$ and the conditional intensity $\lambda (x_o| \Phi_{W_{obs}}) $. \resub{Second, we want to analyze the sensitivity of the predictions when both the intensity and the pair correlation function are unknown, as it is the case in practice.}

\resub{\subsection{Goodness of prediction}}

In order to see how the prediction estimator behaves with respect to the pattern structure we use the inhomogeneous Mat\'ern cluster model which allows the computation of its conditional intensity.
This model is obtained as follows. We first define a stationary Mat\'ern cluster process $\Phi^{Mat}$, i.e. a process where each point of a Poisson parent process is replaced by a \resub{Poissonnian} cluster of offspring uniformly distributed in a disc of radius $R$ around the parent point.
We then focus on the independent $p(x)-$thinned process, where $p(x)$ is a deterministic function on $\bR^2$ with $0 \leq p(x) \leq 1$.  Every point $x$
belonging to $\Phi^{Mat}$  is deleted with probability $1-p(x)$, and again its deletion is independent of locations and possible deletions of any other points \citep{chiu2013}.

Let $\Phi$ be the process of the thinned offspring and $\Phi^{p}$ the parent process (Poisson process with intensity $\kappa$).
The process $\Phi$ is second-order intensity reweighted stationary with intensity $\lambda(x) = \kappa \mu p(x)$, with $\mu$ the expected number
of offspring per parent. The pair correlation function is the one of $\Phi^{Mat}$:
\begin{center}
$g(r) = 1 + \dfrac{2}{\kappa (\pi R)^2} \left( \arccos \left( \dfrac{r}{2R}\right) - \dfrac{r}{2R} \sqrt{1 - \dfrac{r^2}{4R^2}} \right)$, if $0 < r < 2R$, and $g(r) = 1$ otherwise.
\end{center}
According to \cite{gabriel2021}, the local intensity of the thinned Mat\'ern cluster process $\Phi$ is
\begin{equation}  \label{eq:cond}
   \lambda (x_o |  \Phi_{W_{obs}}) = \frac{\mu p(x_o)}{\pi R^2} \int_{b(x_o,R) \cap ({W_{obs}} \cup \dW)} \lambda^p (y | \Phi_{W_{obs}}) \dd y + \kappa \mu p(x_o) \nu \left( b(x_o,R) \backslash (W_{obs} \cup \dW) \right),
\end{equation}
where $\dW$ stands for the outside border of thickness $R$ of the observation window, $b(x_o,R)$ the disc of center $x_o$ and radius $R$, and $\lambda^p (y | \Phi_{W_{obs}})$ is the conditional intensity of parents in ${W_{obs}} \cup \dW$ given the realization of offspring in $W_{obs}$.
This intensity is approximated by
\begin{equation*} 
  \widehat{\lambda^p} (y |  \Phi_{W_{obs}}) =  \frac{1}{\mu p(y) \pi R^2} \sum_{x \in \Phi_{W_{obs}}} \un_{b(x,R)} (y) + \kappa \exp \left( - \frac{\mu}{\pi R^2} \int_{b(y,R) \cap W_{obs}} p(z) \dd z \right),
\end{equation*}
where the first term is the empirical intensity of parents given the observed offspring, and the second term is the intensity of parents with unobserved offspring.
See \cite{gabriel2021} for the approximation of the conditional distribution of parent points given the offspring points and its validation.

\begin{figure}[h!]
  \centering
  \epsfig{file=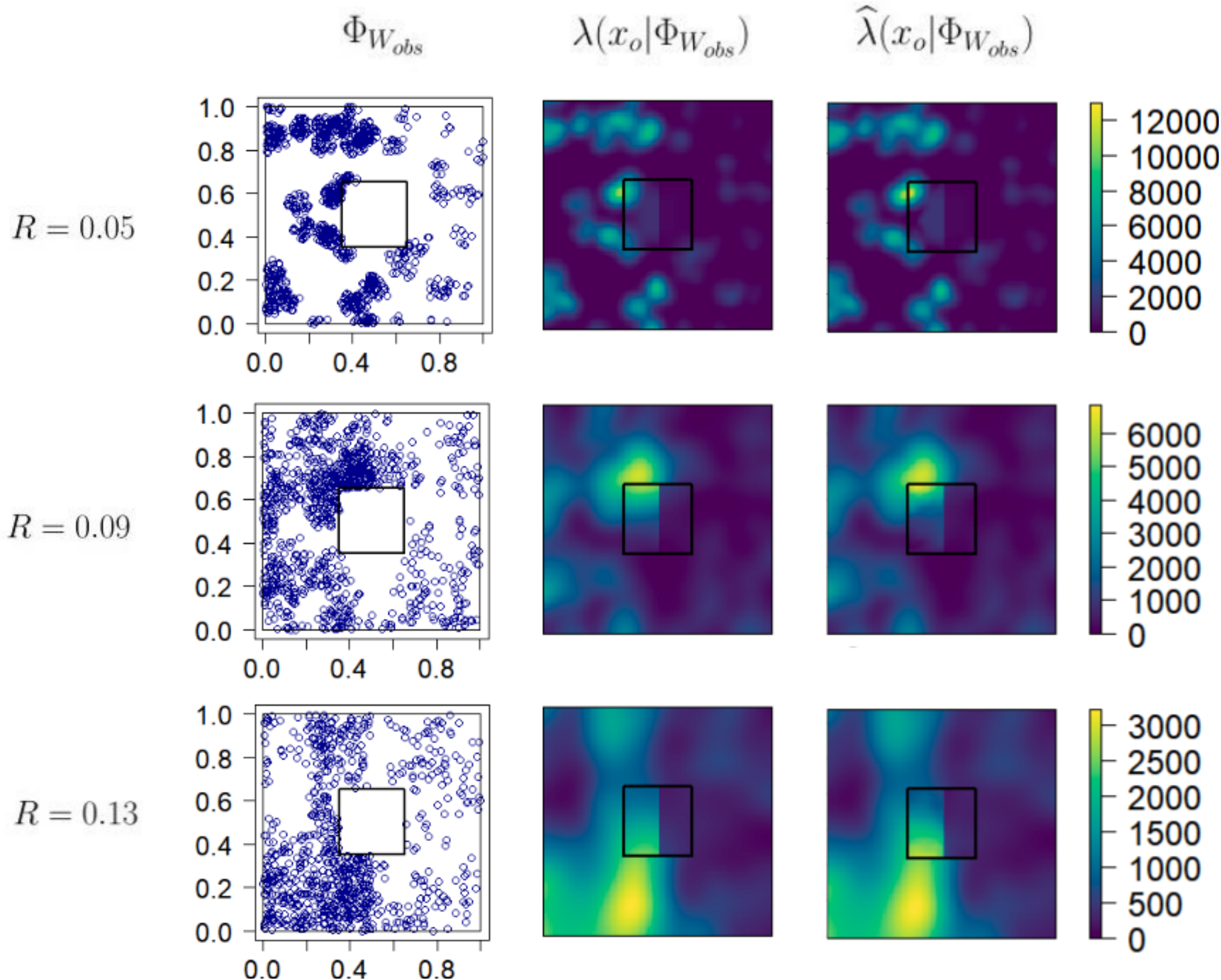,width=0.8\textwidth}
  \caption{First column: inhomogeneous Mat{\'e}rn cluster point patterns, $\Phi_{W_{obs}}$, obtained from $p_1(x)$. Second column: approximation of the local intensity in $W_{pred}$ (central square) and kernel smoothing of the pattern in $W_{obs}$. Third column: predicted local intensity in $W_{pred}$ (central square) and kernel smoothing of the pattern in $W_{obs}$. }\label{fig:patterns}
\end{figure}

\begin{figure}[h!]
  \centering
  \epsfig{file=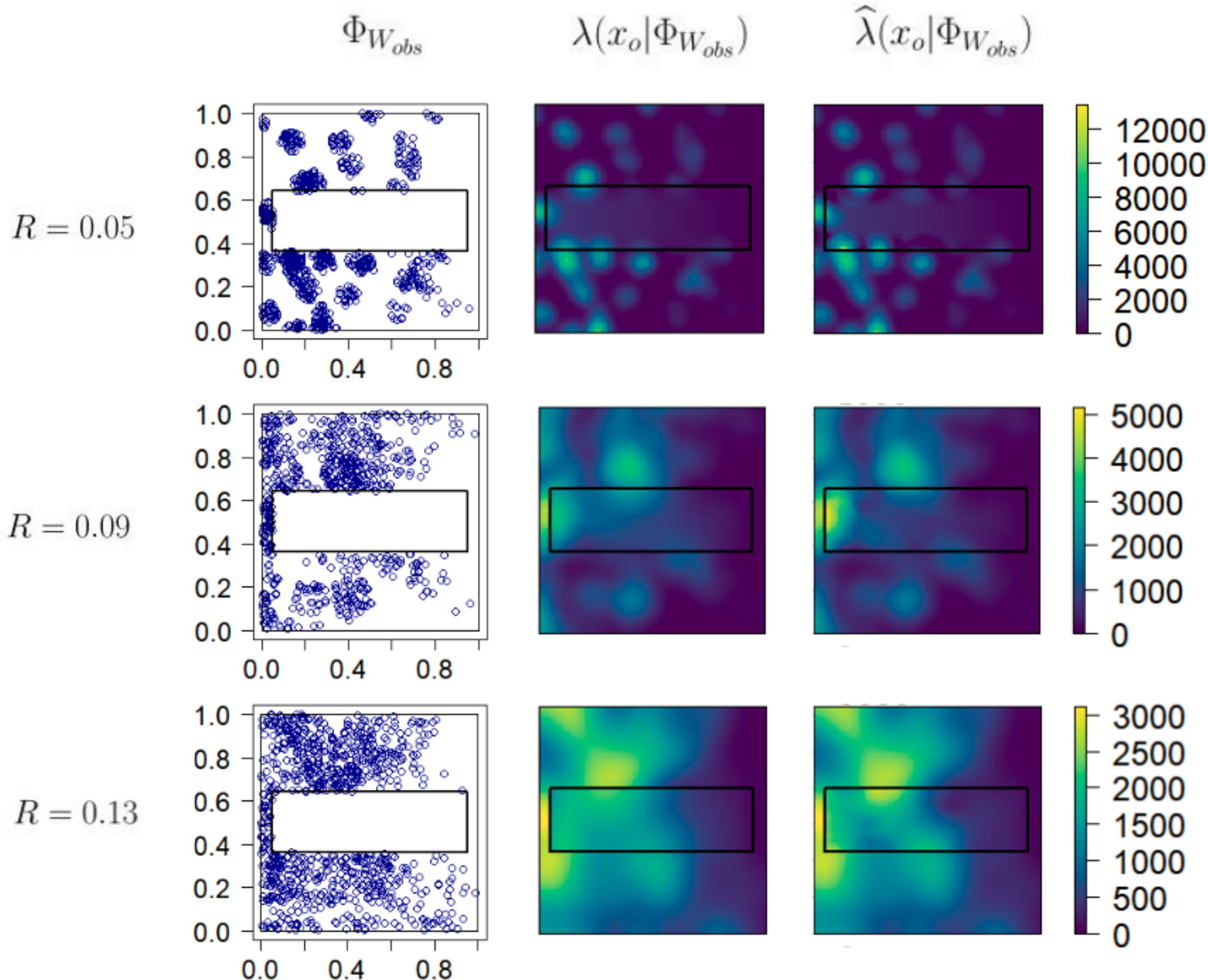,width=0.8\textwidth}
    \caption{First column: inhomogeneous Mat{\'e}rn cluster point patterns, $\Phi_{W_{obs}}$, obtained from $p_2(x)$. Second column: approximation of the local intensity in $W_{pred}$ (central rectangle) and kernel smoothing of the pattern in $W_{obs}$. Third column: predicted local intensity in $W_{pred}$ (central rectangle) and kernel smoothing of the pattern in $W_{obs}$. }\label{fig:patternsH}
\end{figure}

The inhomogeneous Mat{\'e}rn Cluster process (IMCP) $\Phi$ depends on four parameters: the thinning probability $p(x)$, the intensity of parents $\kappa$, the mean number of points per parent $\mu$, and the radius of dispersion of the offspring around the parent points $R$.
In our simulation study we fix $\kappa=50$ and $\mu=40$, and we consider:
\begin{itemize}
  \item $R \in \lbrace 0.05, 0.09, 0.13\rbrace$.

  \item two thinning probabilities:
  $p_1(x)=p_1(x_1,x_2)=\alpha_1 \un_{\lce x_1 \leq v \rce} + \alpha_2 \un_{\lce x_1 > v \rce}$, setting $\alpha_1 = 0.8$, $\alpha_2=0.2$ and $v=0.5$,
  and $p_2(x) = p_2(x_1,x_2) =  1-x_1$.

    \item the unit square as study region $W$. The observation window is $W_{obs} = W \backslash W_{pred}$, where $W_{pred} = \lbrack 0.35, 0.65 \rbrack^2$ when using $p_1(x)$ and
        $W_{pred} = \lbrack 0.05, 0.95 \rbrack \times \lbrack 0.36, 0.64 \rbrack  $ when using $p_2(x)$.
    \end{itemize}
\resub{The size and shape of the prediction windows $W_{pred}$ have been designed to emphasize any effect of the thinning probability on the predictions.}
We perform $n=1000$ simulations of $\Phi$ for all combinations of pairs $(p(x), R)$. Each scenario is denoted IMCP$(p(x), R)$.
Realizations of IMCP$(p_1(x), R)$ (resp. IMCP$(p_2(x), R)$) are given in the first column of Figure~\ref{fig:patterns} (resp. Figure~\ref{fig:patternsH}) for the different values of $R$, illustrating inhomogeneous patterns with increasing range of clustering from top to bottom.
For each simulation, we compute the local intensity from both the predictor (\ref{weightseq}) and the approximation (\ref{eq:cond}) in $W_{pred}$.
We consider the same mesh for the Galerkin approximation method for all configurations, for which $W_{obs}$ is subdivided in 15,194 triangles.
The local intensity and the prediction are respectively plotted in the second and third columns of Figures~\ref{fig:patterns} and \ref{fig:patternsH}.
For a visualization purpose, we plotted the local intensity in $W_{pred}$ (central square/rectangle), as well as a Gaussian kernel smoothing of the  intensity in $W_{obs}$. \resub{Zoom of the local intensity and the predictions in $W_{pred}$ are given in Figures~\ref{fig:zoompred} and \ref{fig:zoompredH} (see Appendix~\ref{app:zoom}).}

We can see that the method reproduces the structures of the point process.
In particular, it reproduces clusters, as soon as there are points close enough to the boundary of $W_{pred}$, and the prediction tends to the intensity of the process, $\lambda(x) = \kappa \mu p(x)$, when it is made at distances larger than $R$ from the boundary of $W_{obs}$. This is particularly clear for small values of $R$ (first rows in Figures~\ref{fig:patterns} and \ref{fig:patternsH}).
Results from these realizations show, however, some weak differences between the local intensity and our prediction that we  now quantify.

\begin{figure}[h!]
\centering
\begin{subfigure}{.45\textwidth}
  \centering
  \includegraphics[width=\linewidth]{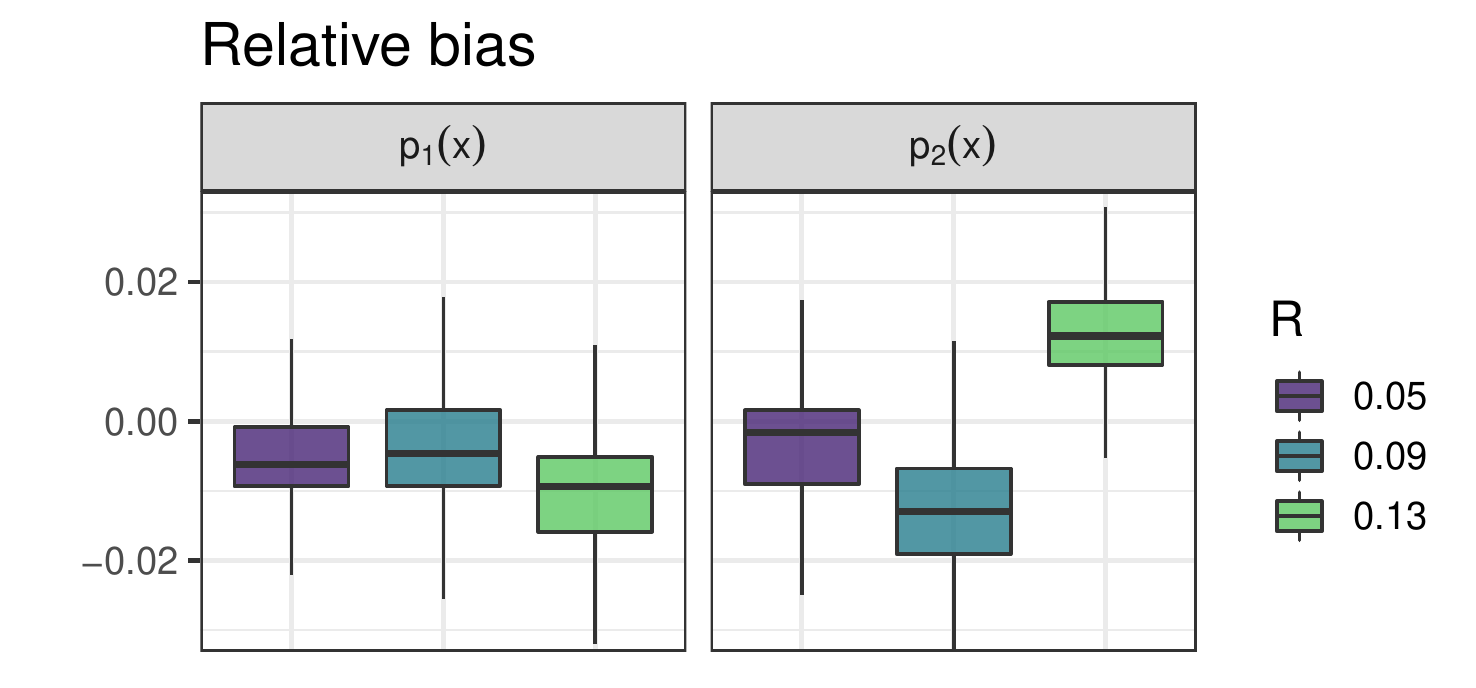}
  \vspace{1.5mm}
  \caption{}\label{3:a}
\end{subfigure}%
\begin{subfigure}{.45\textwidth}
  \centering
  \includegraphics[width=\linewidth]{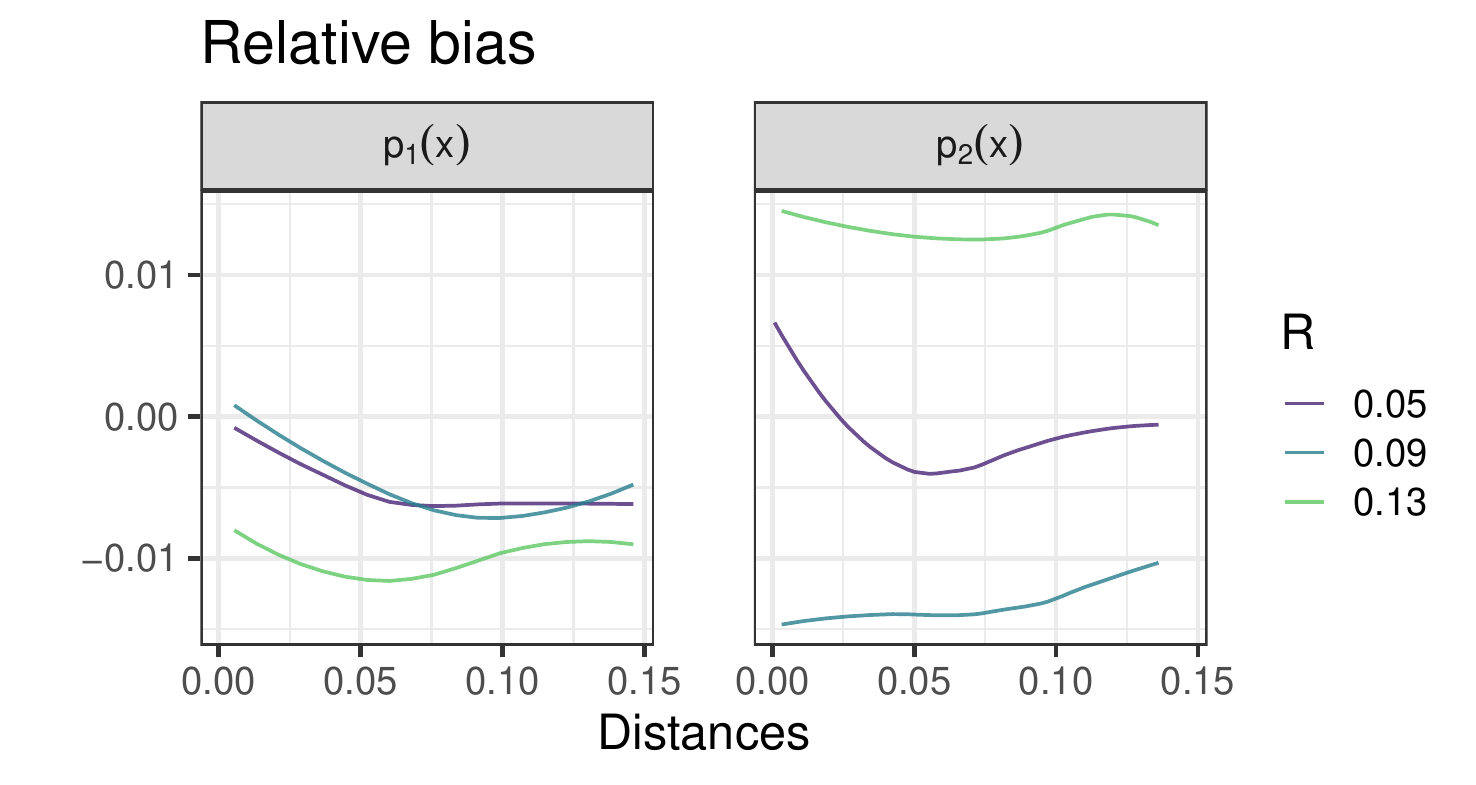}
    \caption{}\label{3:b}
\end{subfigure}

\begin{subfigure}{.45\textwidth}
  \centering
  \includegraphics[width=\linewidth]{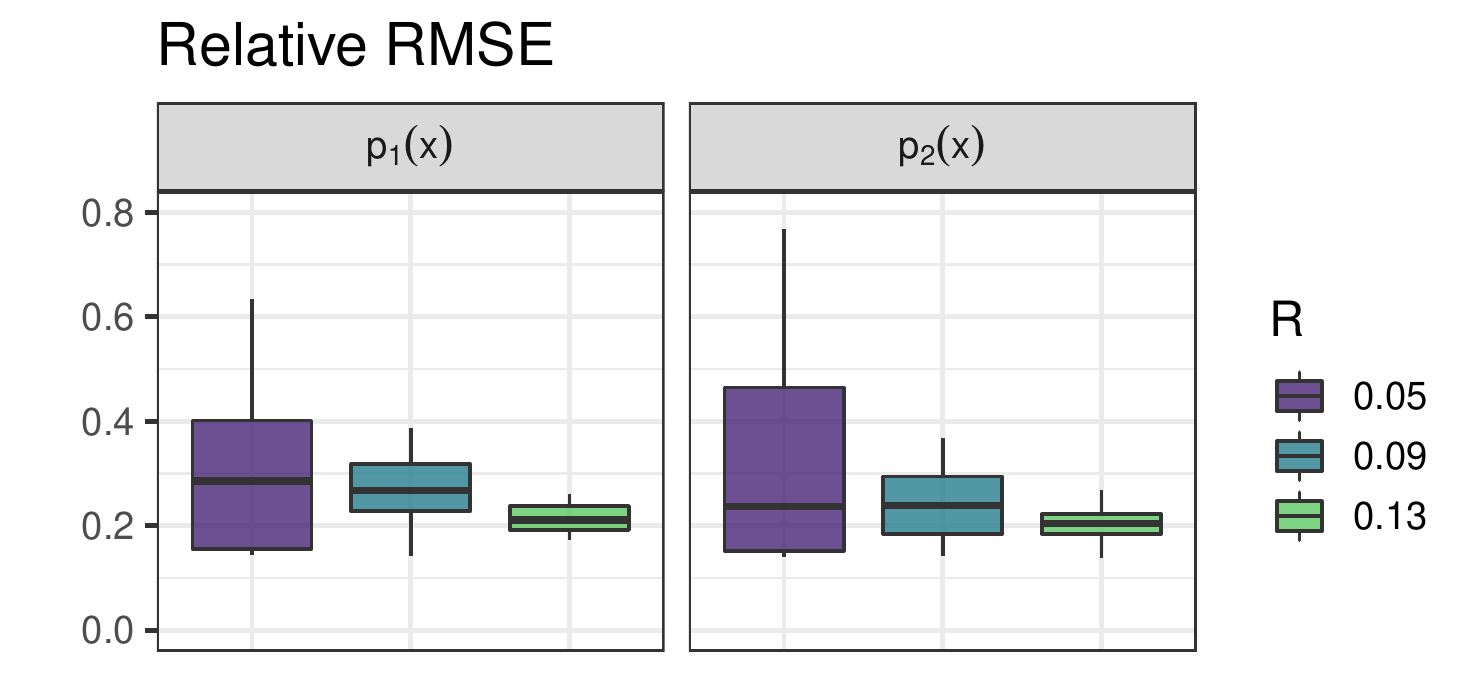}
    \vspace{1.5mm}
    \caption{}\label{3:c}
\end{subfigure}
\begin{subfigure}{.45\textwidth}
  \centering
  \includegraphics[width=\linewidth]{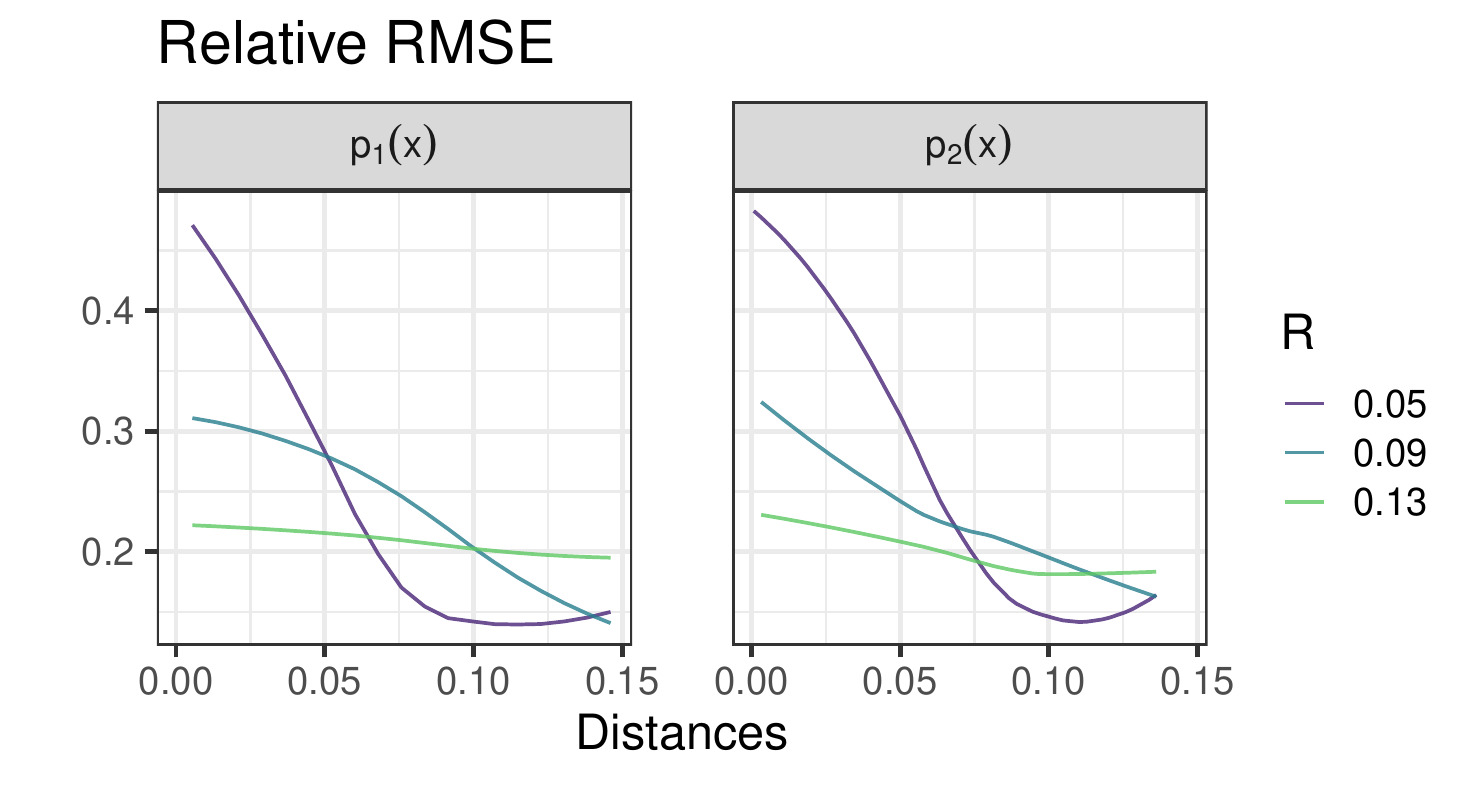}
    \caption{}\label{3:d}
\end{subfigure}
\caption{Relative bias (top) and relative RMSE (bottom) computed from 1000 simulations with $p_1(x)$ and $p_2(x)$ within $W_{pred}$. Distances are measured outwards the boundary of $W_{obs}$, inwards $W_{pred}$.}\label{fig:relativeBMSE}
\end{figure}

We measure the precision and the variability of our prediction through the relative bias (RB) and the relative root mean square error (RRMSE) as follows
\begin{equation*}
\mbox{RB}(x_o; \hat \lambda, \lambda) = \frac{\sum\limits_{i=1}^n \left( \widehat{\lambda_i}(x_o|\Phi_{W_{obs}}) - \lambda_i (x_o|\Phi_{W_{obs}}) \right)}{\sum\limits_{i=1}^n  \lambda_i (x_o|\Phi_{W_{obs}})},
\end{equation*}
\begin{equation*}
\mbox{RRMSE}(x_o; \hat \lambda, \lambda) = \frac{\sqrt{\frac{1}{n}\sum\limits_{i=1}^n \left( \widehat{\lambda_i}(x_o|\Phi_{W_{obs}}) - \lambda_i (x_o|\Phi_{W_{obs}}) \right)^2 }}{ \frac{1}{n}\sum\limits_{i=1}^n  \lambda_i (x_o|\Phi_{W_{obs}})},
\end{equation*}
where $\widehat{\lambda_i}$ and $\lambda_i$ stand for the prediction and the local intensity of the $i$-th simulation.
\resub{We compute RB$(x_o)$ and RRMSE$(x_o)$ at all $x_o \in W_{pred}$. This allows us to get an insight of their distribution as illustrated in Figure~\ref{3:a} and~\ref{3:c} for all patterns IMCP$(p(x),R)$. We also look over the values of RB$(x_o)$ and RRMSE$(x_o)$ according to the distance $d(x_o, W_{pred})$ between the point $x_o$ where the prediction is made and the boundary of the prediction window $W_{pred}$, as plotted in Figure~\ref{3:b} and~\ref{3:d} for all patterns IMCP$(p(x),R)$.}

\begin{figure}[h!]
  \centering
  \includegraphics[width=.4\linewidth]{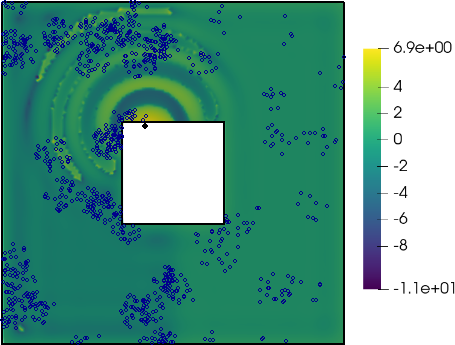}
    \caption{Weight function $w(x;x_o)$ (here transformed as $\text{sign}(w(x;x_o)) \log ( |w(x;x_o)| )$) associated to the point $x_o$ (black dot in $W_{pred}$) for the process IMCP$(p_1(x),0.05)$. For each realization of the point process, $\Phi_{W_{obs}}$ (e.g. the dark blue dots), the local intensity at $x_o$ is the sum over $x \in \Phi_{W_{obs}}$ of the weights $w(x;x_o)$.  }  \label{fig:poids}
\end{figure}

In all cases, the relative bias is concentrated around 0 and is much less than 5\% (in absolute value) \resub{which reveals} an excellent precision of the predictor.
The distributions of RRMSE$(x_o)$ show a decreasing variability as the interaction radius $R$ increases.  Indeed, as shown in Figures~\ref{fig:patterns} and~\ref{fig:patternsH}, the inter-point distances of the point patterns are greater for
larger values of $R$ and the local intensity is thus more diffuse for large values of $R$, and more picky for very small values. The weight function is closely related to the pair correlation function. It shows a radial wavy behavior around $x_o$ with positive decreasing values at distances less that $R$, and it has negative values between $R$ and $2R$, \dots,  and then is null (see Figure~\ref{fig:poids}).
As the predictor is the sum of the weights at $x \in \Phi_{W_{obs}}$, it reflects this behavior and clusters are thus well
\resub{identified}.
 The approximation of the local intensity (\ref{eq:cond}) is smoother as it is only expressed in terms of first-order moments. Hence, we get more variability in the border $\dW$, and even higher for picky intensities as for $R=0.05$. That is also shown in Figures~\ref{3:b} and~\ref{3:d} which illustrate the relative bias and relative RMSE computed at different distances from $\dW$. They are both higher at very short distances.

\begin{figure}[h!]
  \centering
   \epsfig{file=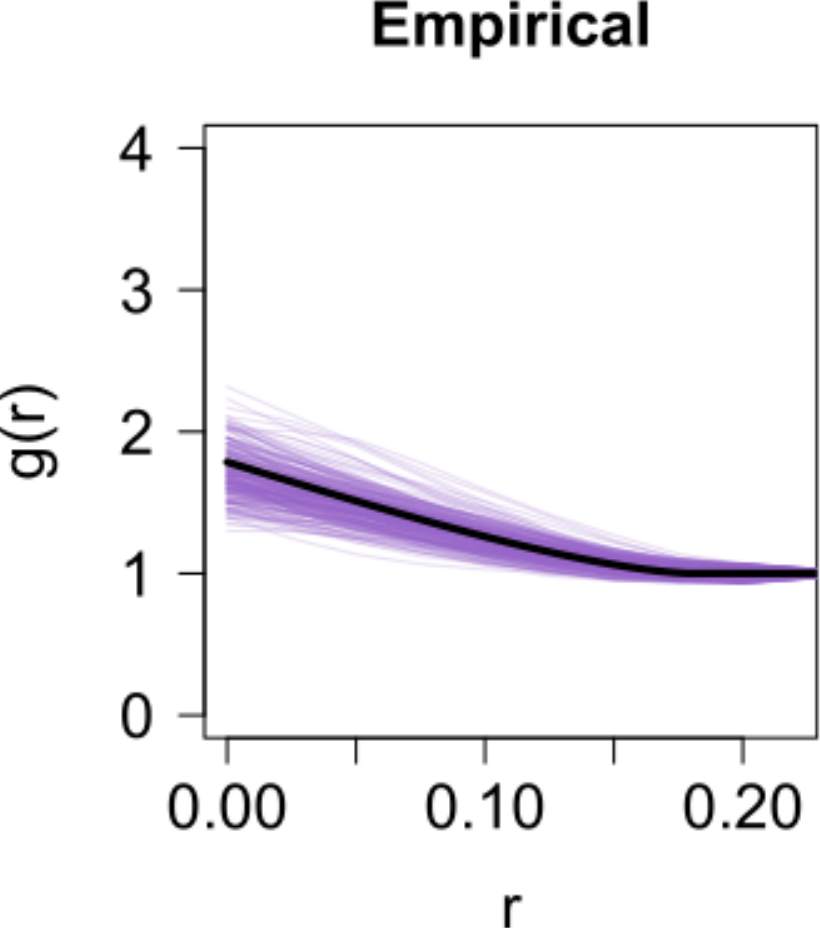,width=0.25\textwidth} \hspace{2mm}
   \epsfig{file=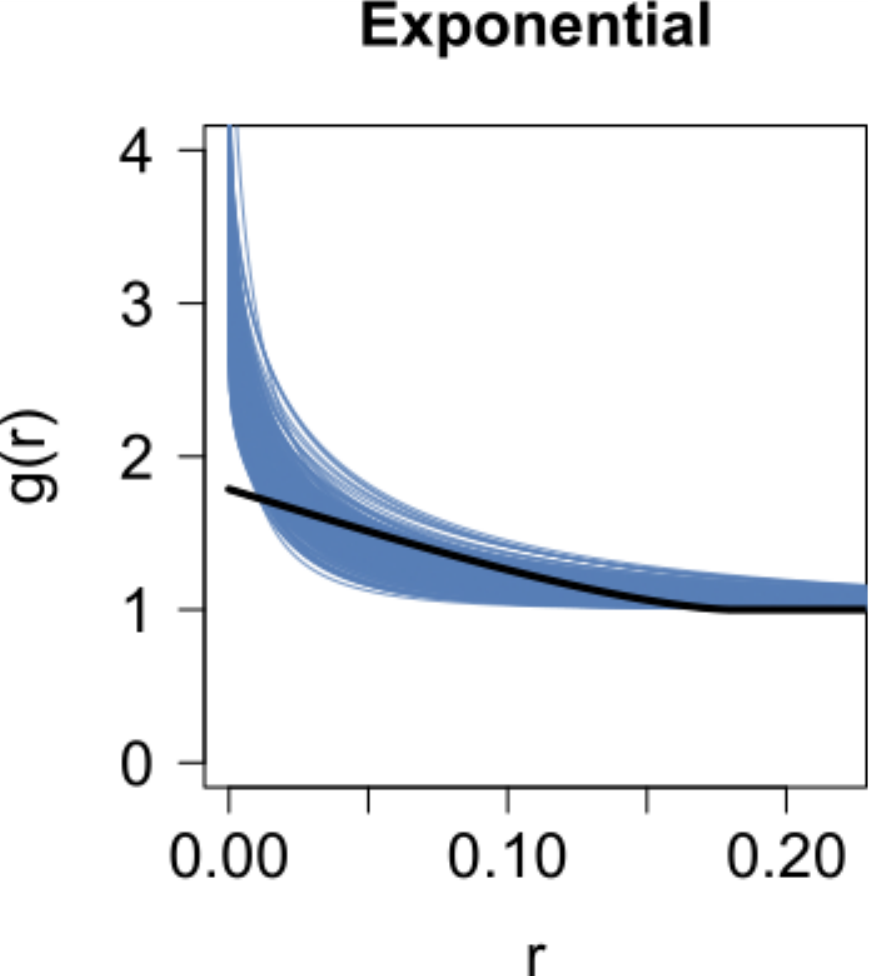,width=0.25\textwidth} \hspace{2mm}
   \epsfig{file=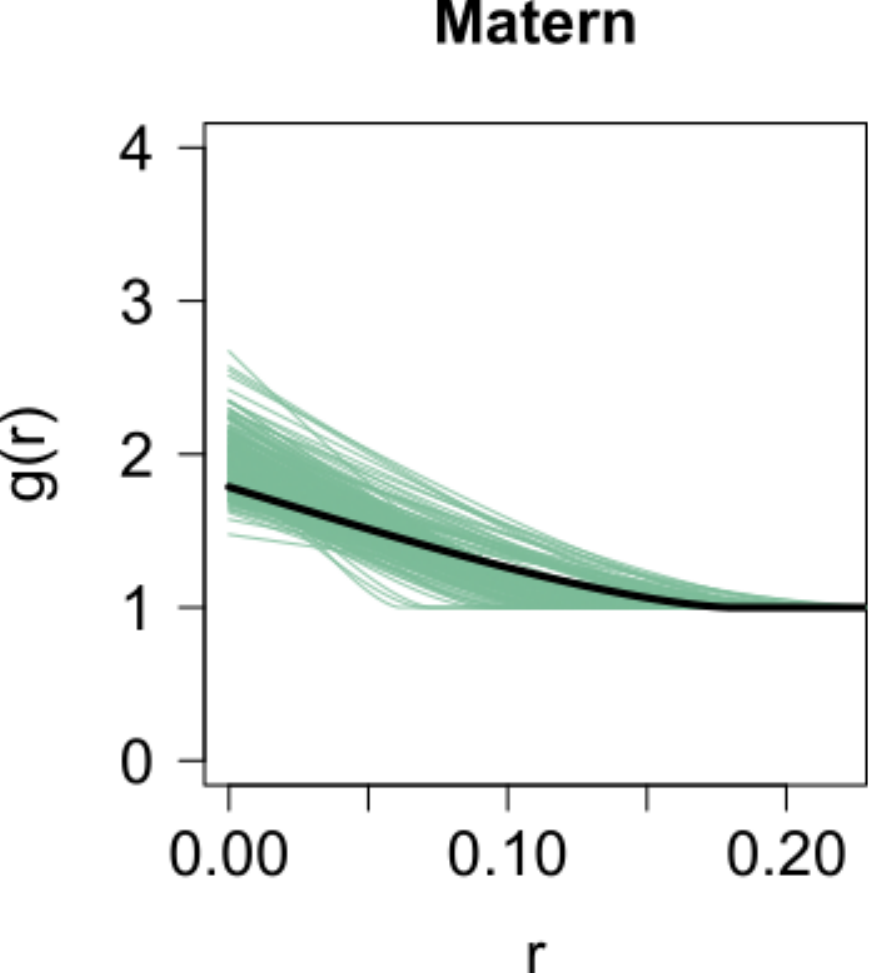,width=0.25\textwidth}
  \caption{Fitted pair correlation functions: empirical (purple), exponential model (blue), Mat{\'e}rn model (green). The black curve is the theoretical pair correlation function.} \label{fig:fittedpcf}
\end{figure}

\resub{\subsection{Sensitivity to the estimation of the first- and second-order moments}
\label{sec:sensitivity}

In the previous section the intensity and the pair correlation functions of the inhomogeneous Mat{\'e}rn process were assumed to be known. We now aim to measure the effect of their estimation on the predictions. We focus on the process IMCP$(p_1(x),0.09)$ and we consider four situations for the prediction:
\begin{itemize}
  \item  using the theoretical intensity and pair correlation functions (as defined in the previous section). The resulting predictions are used as  the basis for further comparisons.
   \item using a parametric model for the intensity function: $f_\lambda (x) = \beta_1 \un_{\lce x_1 \leq 0.5 \rce} + \beta_2 \un_{\lce x_1 > 0.5 \rce}$ and  for the pair correlation function:
  \begin{itemize}
    \item either the Mat{\'e}rn model: $g_{mat}(r) = 1 + \frac{2}{\alpha_1 (\pi \alpha_2)^2} \left( \arccos \left( \frac{r}{2 \alpha_2}\right) - \frac{r}{2 \alpha_2} \sqrt{1 - \frac{r^2}{4 \alpha_2^2}} \right)$, if $0 < r < 2 \alpha_2 $, and $g(r) = 1$ otherwise. This allows us to analyze the effect of fitting on the true model.
    \item or the exponential model: $g_{exp}(r) = 1 + \frac{1}{\alpha_3} \exp \left( - \alpha_4 \sqrt{r} \right)$. This model allows us to check the effect of a misspecification of the pair correlation function.
    \item  or non-parametric estimation, $g_{emp}$, obtained by kernel smoothing, as implemented in the \texttt{pcfinhom} function in \texttt{spatstat} with $f_\lambda (x)$ as intensity function. This situation is often the most natural one and also the one with the less hypotheses. Because empirical estimates do not always satisfy theoretical conditions of a pair correlation function (positive definiteness, consistence), we made an \textit{a posteriori} selection of the admissible fitted pair correlation functions (see Appendix~\ref{app:estimation}).
  \end{itemize}
\end{itemize}
All results are presented for 250 admissible simulations.
We consider a maximum likelihood estimator of the intensity of a Poisson point process to estimate parameters $\beta_1$ and $\beta_2$, i.e.
\begin{center}
$\hat \beta_1 = \Phi\left( W_{obs} \cap [0,0.5]\times [0,1] \right) / \nu\left( W_{obs} \cap [0,0.5]\times [0,1] \right)$ and $\hat \beta_2 = \Phi\left( W_{obs} \cap [0.5,1]\times [0,1] \right) / \nu\left( W_{obs} \cap [0.5,1]\times [0,1] \right)$.
\end{center}
Parameters $\alpha_i$, $i=1,\dots,4$ are estimated by non-linear least squares. Fitted pair correlation functions are plotted in Figure~\ref{fig:fittedpcf}. Summaries of fitted parameters are postponed to  Appendix~\ref{app:estimation} (Table~\ref{table:param}).}

\resub{
To compare the predictions we compute $\mbox{RB}(x_o; \widehat \lambda, \lambda)$ and $\mbox{RRMSE}(x_o; \hat \lambda, \lambda)$, where $\widehat{\lambda}$ denotes the prediction using either the theoretical moments or the fitted ones and $\lambda$ denotes the local intensity~(\ref{eq:cond}).
As in the previous section, the relative bias and relative RSME are computed at all $x_o \in W_{pred}$ so that we can plot their distribution, see Figure~\ref{fig:comppcf}a and~\ref{fig:comppcf}b. These figures show that the relative bias  is about one tenth of the relative RMSE. The relative bias tends to be positive, slightly larger when the predictions are made from $(f_\lambda, \ g_{exp})$. The relative RMSE are in the same order of magnitude, slightly smaller for $(f_\lambda, \ g_{exp})$.
\begin{figure}[h!]
  \centering
  \begin{subfigure}{0.45\textwidth}
  \centering
   \epsfig{file=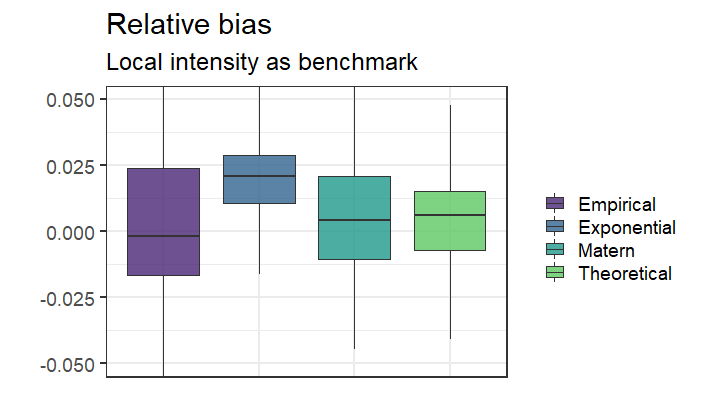,width=0.95\textwidth}
    \caption{}\label{compcf:a}
\end{subfigure}
  \begin{subfigure}{0.45\textwidth}
  \centering
     \epsfig{file=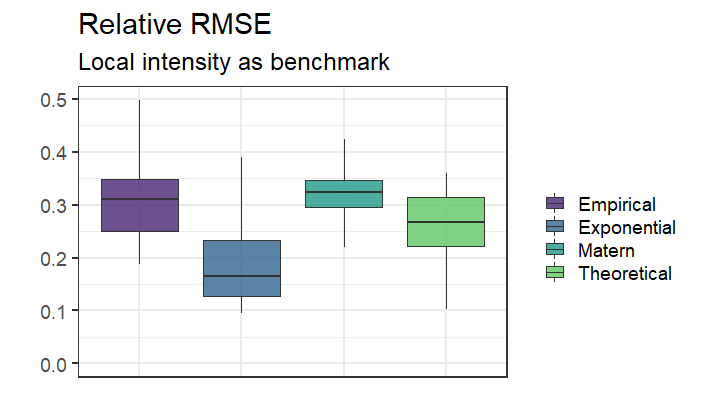,width=0.95\textwidth}
    \caption{}\label{compcf:b}
\end{subfigure}

    \begin{subfigure}{0.45\textwidth}
  \centering
   \epsfig{file=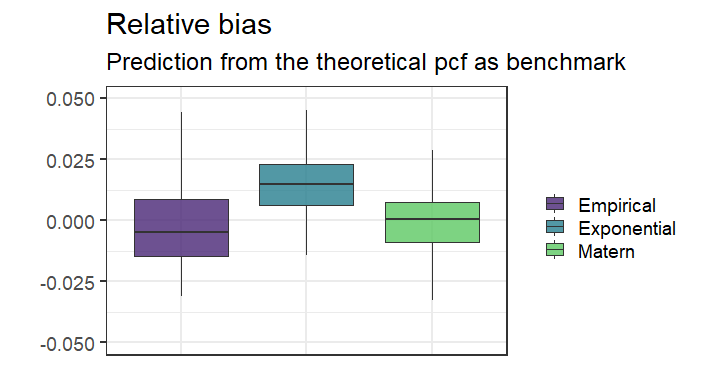,width=0.95\textwidth}
         \caption{}\label{compcf:c}
     \end{subfigure}
         \begin{subfigure}{0.45\textwidth}
  \centering
     \epsfig{file=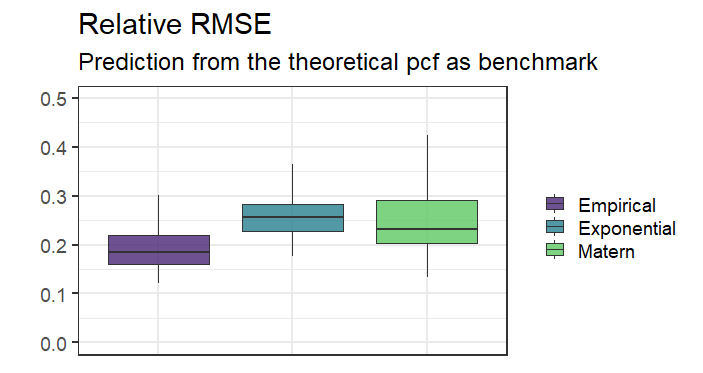,width=0.95\textwidth}
         \caption{}\label{compcf:d}
     \end{subfigure}
  \caption{Relative bias (left) and relative RMSE (right) computed from the local intensity (top) or from the prediction with the theoretical moments (bottom).} \label{fig:comppcf}
\end{figure}

To emphasize the effect of using estimated moments rather than the theoretical ones, we also compute $\mbox{RB}(x_o; \widehat \lambda, \widehat \lambda_{theo})$ and $\mbox{RRMSE}(x_o; \widehat \lambda, \widehat \lambda_{theo})$, where $\widehat{\lambda}$ (resp.  $\widehat \lambda_{theo}$) denotes the prediction using the fitted intensity and pair correlation functions (resp. the theoretical ones). Figure~\ref{fig:comppcf}c and~\ref{fig:comppcf}d illustrate the boxplot of their values over all $x_o \in W_{pred}$.
The ratio between the relative bias and relative RMSe remains the same.
The relative bias is close to zero for both the predictions made using $(f_\lambda, \ g_{emp})$ and $(f_\lambda, \ g_{mat})$ and slightly larger and positive for $(f_\lambda, \ g_{exp})$. The relative RMSE is smaller for $(f_\lambda, \ g_{emp})$.

In the former case, the relative bias and relative RMSE are computed from the local intensity which is a slightly smoothed version of the true conditional intensity, whereas in the later case they are computed from the prediction obtained with the theoretical intensity and pair correlation functions leading to less smoothed predictions. As the exponential model also smoothes the predictions, its relative RMSE in the former case is low, but increases in the later case and become similar to the one of from the Mat{\'e}rn model of the pair correlation function. Both predictions from $(f_\lambda, \ g_{emp})$ and $(f_\lambda, \ g_{mat})$ have similar behavior than the one from the theoretical moments. Figure~\ref{fig:preddpcf} compare the predictions from the different cases on a single simulation and illustrate all these comments.
\begin{figure}[h!]
  \centering
   \epsfig{file=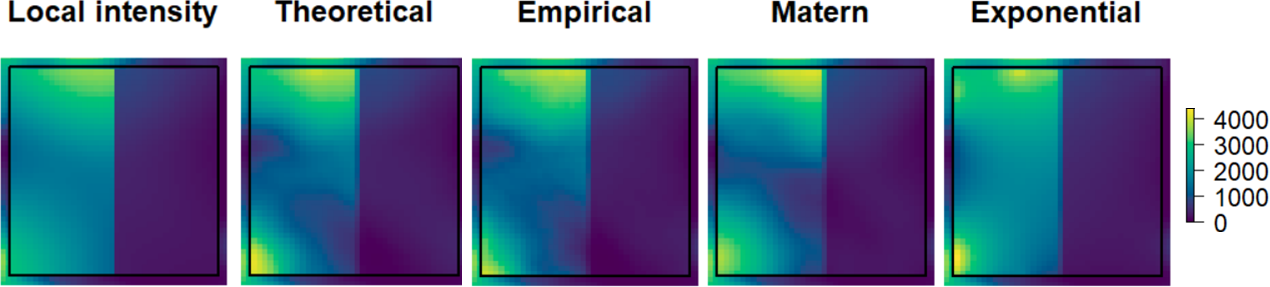,width=0.75\textwidth}
  \caption{Local intensity (left) and predictions obtained from the theoretical moments or their non-parametric (empirical) or parametric (Matern, Exponential) estimation.} \label{fig:preddpcf}
\end{figure}
Whilst the results from the non-parametric estimation of the pair correlation function are promising, they do not systematically rely on an admissible pair correlation function (leading to numerical instabilities in the predictions).
}

\section{Predicting the density of earthquakes}
\label{sec:application}

\begin{figure}[h!]
  \centering
   \epsfig{file=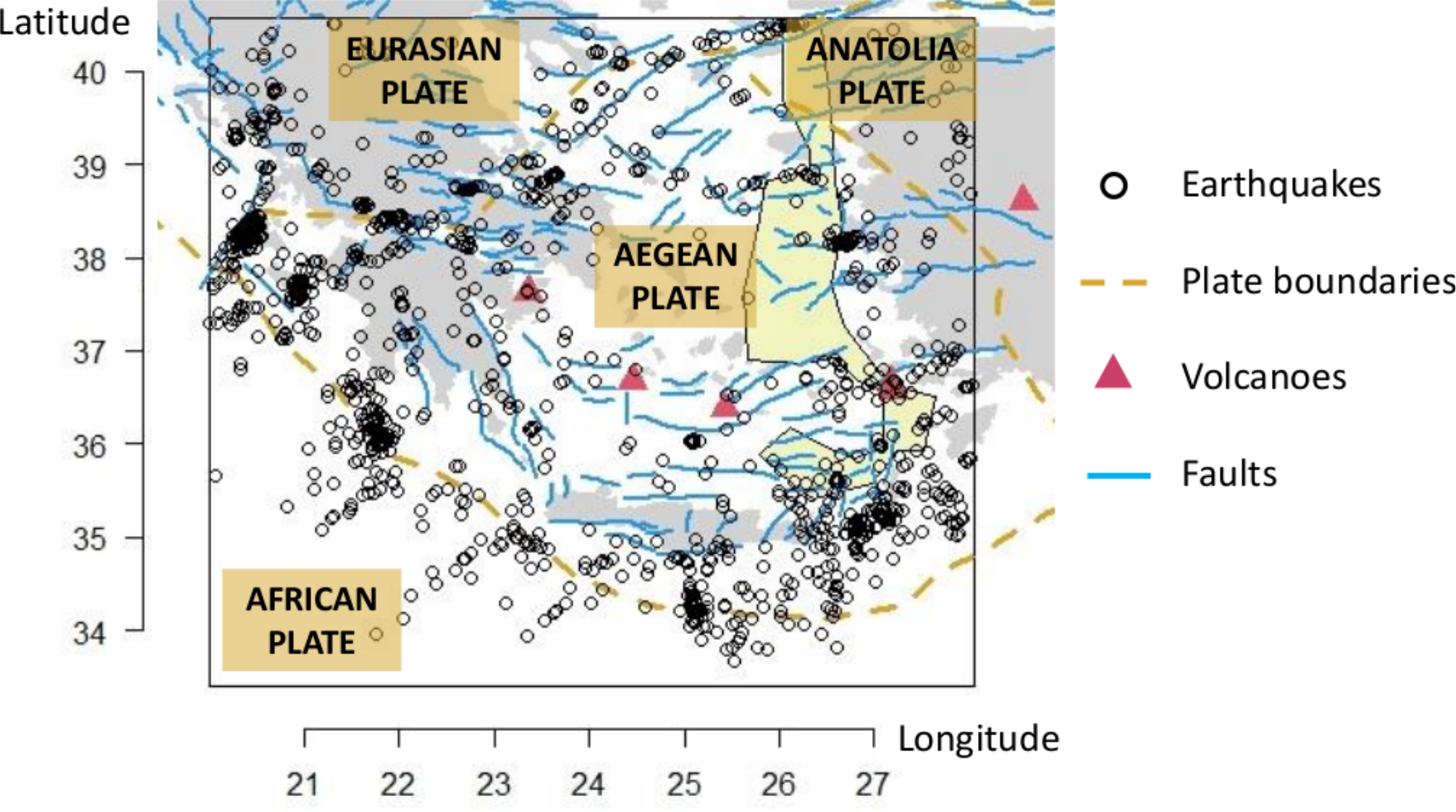,width=0.8\textwidth}
  \caption{Earthquakes of magnitude greater or equal than 4, faults, plate boundaries and main volcanoes in the Hellenic region.  Prediction window is shown in the light yellow area.} \label{fig:patternearthquakes}
\end{figure}

In this section we focus on the Greek-Hellenic area, a region of high seismic hazard due to both tectonic and volcanic seismogenic sources. This seismicity is a result of some motion-induced deformations: northward motion of the African lithosphere, westward motion of the Anatolia plate and collision between African and Eurasian plates
\citep{papazachos1971,lepichon1979,mckenzie1972,anderson1987}. A total of 1173 earthquakes of magnitude greater or equal than 4 occurred in the study area $W$ (black square in Figure~\ref{fig:patternearthquakes}) between 2004 and 2015. Their locations are plotted in Figure~\ref{fig:patternearthquakes}.
Data have been recorded by the Hellenic Unified Seismological Network (HUSN).
Seismic networks provide data that are used as the basis both for public safety decisions and for scientific research. They are essential tools for observing earthquakes and assessing seismic hazards that can be described and characterized to assess their degree of coverage \citep{siino2020}.
Indeed, their configuration affects the data completeness, which in turn, critically affects several seismological scientific targets (e.g., earthquake prediction, seismic hazard, etc. \citep{vamvakaris2013}). From different indicators (magnitude of completeness, Radius of Equivalent Sphere that estimates an average error of location), \cite{dalessandro2011} identified some seismogenic areas that probably are not adequately covered by the HUSN. Based on their results we delineate a zone (in light yellow in Figure~\ref{fig:patternearthquakes}) with unreliable or missing records. We removed the 48 points located in this yellow zone to predict the local intensity of earthquakes.

\begin{figure}[h]
  \centering
   \epsfig{file=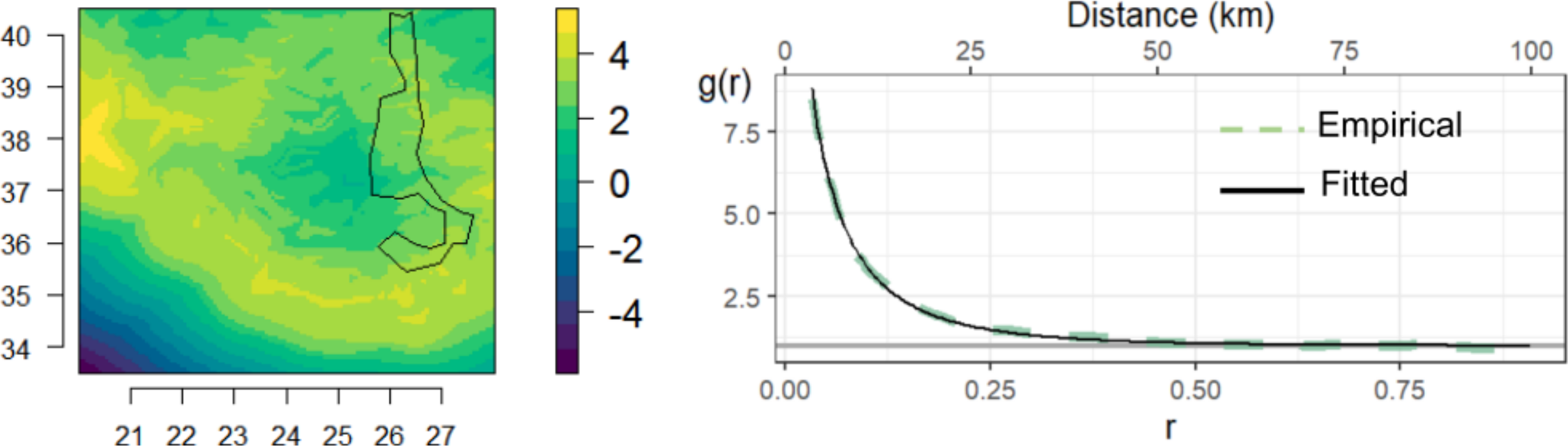,width=\textwidth}
  \caption{Log-intensity $\hat \lambda(x)$ (left panel) and pair correlation function (right panel): empirical (green dashed line), fitted (black solid line).
  }\label{fig:intpcfearthquakes}
\end{figure}

Modelling earthquakes needs to account for geological information and interactions between earthquakes, usually in terms of clustering. The most popular models of seismological events are the so-called self-exciting, such as Hawkes and ETAS models \citep{adamopoulos1976,ogata1988}. Our approach, however, is model-free, it accounts for environmental heterogeneity and interaction between events.
Following \cite{siino2017}, we assume that the intensity of earthquakes log-linearly depends on several spatial geological covariates: $D_{pb}$, the distance to the plate boundary (orange dashed curves in Figure~\ref{fig:patternearthquakes}); $D_f$, the distance to the nearest fault (blue lines in Figure~\ref{fig:patternearthquakes}); and $D_v$, the distance to the nearest main volcano (red triangles in Figure~\ref{fig:patternearthquakes}). Thus, the intensity takes the form
\begin{multline}
\lambda(x) = \lambda(x_1,x_2) = \exp \left( \beta_0 + \beta_1 x_1 + \beta_2 x_2 + \beta_3 x_1^2 + \beta_4 x_1 x_2 + \beta_5 x_2^2 + \beta_6  \un_{\lce D_f(x) \leq \phi_1 \rce} D_f(x)
\right. \\
\left.  + \beta_7  \un_{\lce \phi_1 < D_f(x) \leq \phi_2 \rce} D_f(x) + \beta_8 \un_{\lce \phi_2 < D_f(x) \leq \phi_3 \rce} D_f(x) + \beta_9 \un_{\lce \phi_3 < D_f(x) \leq \phi_4 \rce} D_f(x)
\right. \\
\left.
+ \beta_{10} \un_{\{D_f(x) > \phi_4\}} D_f(x)  + \beta_{11} D_v(x) + \beta_{12} D_{pb} (x)
  \right),
  \label{eq:intearthquakes}
\end{multline}
with $\phi_1=6.73$, $\phi_2=43.48$, $\phi_3=54.783$ and $\phi_4=112$.
The estimated values of model parameters are \resub{obtained by the method of maximum likelihood from all points $W_{obs}$ and are} given in Table~\ref{table:betas}, and the estimated intensity is depicted in Figure~\ref{fig:intpcfearthquakes} (left panel, log-scaled).
This figure shows that the region to be predicted (bordering in black) is across zones of strong and weak intensities.
\begin{table}[h]
\centering
\caption{Estimated parameters of the intensity model $\lambda(x)$.}
\scalebox{1.0}{
\begin{tabular}{c|rrrrrrr}
  \hline
  Parameter & $\beta_{0}$ & $\beta_{1}$ & $\beta_{2}$ & $\beta_{3}$ & $\beta_{4}$ & $\beta_{5}$ & $\beta_{6}$  \\   \hline
  Estimate & -178.483    &   0.611     &  9.673   &    0.034    &  -0.061 &     -0.114  &    -0.034  \\
  \hline   \hline
    Parameter &  $\beta_{7}$ & $\beta_{8}$ & $\beta_{9}$ & $\beta_{10}$ & $\beta_{11}$ & $\beta_{12}$ & \\   \hline
  Estimate & -0.0268   &  -0.0175   &  -0.0179  &   -0.0214 &     0.0040   &  -0.0093 & \\
  \hline
\end{tabular}}
\label{table:betas}
\end{table}
We then calculated the empirical pair correlation function
\resub{using kernel methods and the parametric estimate of the intensity
 (see Figure~\ref{fig:intpcfearthquakes} (right panel)). We then} fitted
an exponential model of the form
$g(r) = \exp(- \alpha_2 \sqrt{r})/\alpha_1$, with $\hat \alpha_1=8.9502$ and $\hat \alpha_2=0.0266$ (as it is shown in Figure~\ref{fig:intpcfearthquakes} (right panel), for the theoretical fitted model). \resub{Model parameters are estimated as in section~\ref{sec:sensitivity}.}
The high values of the pair correlation function at short distances are indicative of a cluster process with a small range, less than $50~km$. Then we have $g(r) = 1$ (horizontal dark grey line), indicating there is no particular spatial structure or interaction. Note that $g(0) > 7.5$ corresponds to a high probability of observing an earthquake within a small disc centered at an observed earthquake.

\begin{figure}[h]
  \hspace{1.5cm}$\widehat \lambda(x_o | \Phi_{w_{obs}})$ \hspace{3cm} $\widehat \lambda(x_o | \Phi_{w_{obs}})$ \hspace{8mm} $\widehat \lambda(x_o)$ \hspace{3cm} $\widehat \lambda(x_o | \Phi_{w_{obs}}) / \widehat \lambda(x_o)$

    \centering
        \epsfig{file=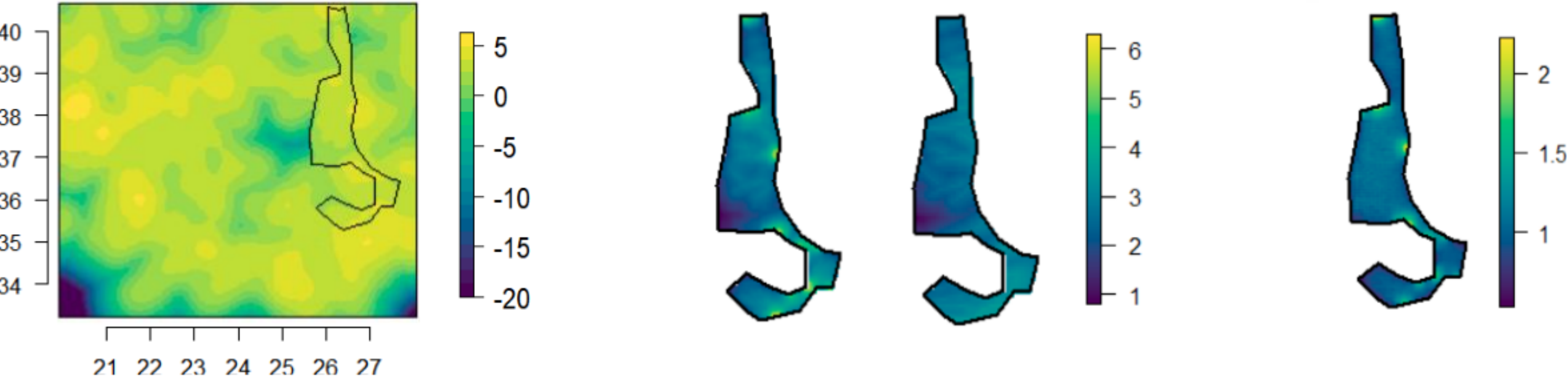,width=\textwidth}
  \caption{ Prediction of the local intensity $\widehat \lambda(x_o| \Phi_{W_{obs}})$ in $W_{pred}$, and a Gaussian kernel smoothing in $W_{obs}$ (left panel).
  Prediction of the local intensity and fitted intensity $\widehat \lambda(x_o)$ given in Equation~(\ref{eq:intearthquakes}) in $W_{pred}$ (middle panel) and their ratio (right panel).
  All plots are at a log-scale.}\label{fig:predictionearthquakes}
\end{figure}

The prediction of the local intensity is obtained by using a mesh
of 22,686 triangles for the Galerkin approximation method.
The left panel of Figure~\ref{fig:predictionearthquakes} shows the predicted local intensity  $\widehat \lambda(x_o| \Phi_{W_{obs}})$ in $W_{pred}=W \backslash W_{obs}$
 and a Gaussian-kernel smoothing of observed earthquake locations in $W_{obs}$, with bandwidth $22~km$. Both intensities are at a log-scale. The blue area in the center
of $W_{obs}$ indicates a kernel smoothing value near zero because of the absence of points in this region (see Figure~\ref{fig:patternearthquakes}).
Compared with Figure~\ref{fig:intpcfearthquakes}, this plot also emphasizes the differences between the intensity fitted from covariates in Equation~(\ref{eq:intearthquakes}) and the empirical intensity obtained by kernel smoothing.
This leads to conclude that we can expect some influence from point locations on the local intensity. This effect is observed in the middle panel of Figure~\ref{fig:predictionearthquakes}. It focuses on the prediction window in which both  the predicted local intensity
and the fitted intensity of the point pattern $\widehat \lambda(x_o)$  are plotted at a log-scale. The right panel plots their ratio to highlight their differences.
The different range values between $\widehat \lambda(x_o| \Phi_{W_{obs}})$ and
$\widehat \lambda(x_o)$ indicate that if we did not take into account the effect of earthquakes locations in $W_{obs}$, we would get regular variations of the local intensity in $W_{pred}$. The knowledge of these points add local hot spots \resub{close to} the border between $W_{obs}$ and $W_{pred}$.
Hence, the prediction window shows a high and spatially varying local intensity, whose variations reflect the scales of the structures of the underlying process, through  both the intensity and the pair correlation function.

Note that from a practical point of view, the approximation of the weight functions has been implemented in \texttt{FreeFem++}~\citep{hecht2012} a high level integrated development environment for numerically solving partial differential equations. All other computations are implemented in \texttt{R}~\citep{R}. We used the  \texttt{R} package  \texttt{spatstat}~\citep{baddeley2005} for fitting the intensity $\lambda(x)$ and for computing the empirical pair correlation function. Because interfacing \texttt{R} and \texttt{FreeFem++} is not straightforward and not feasible on any operating system, codes are only available upon request to the first author.

\section{Discussion}
\label{sec:discussion}
We have proposed a predictor of the local intensity of point processes conditionally to some censored observations, and accounting for the individual relationships and considering environmental covariates. In this general context we can distinguish two situations: in a first case, censoring is a continuous function defined over the study region $W$, so that observed and unobserved points share the same space; in a second case, observations are taken in some windows and we want to predict outside the observation window. In this paper we presented the latter case and considered that the spatial variation driving the point density is related to an intensity governed by environmental covariates.

When observed and unobserved points share the same space, the study region is the observation window and the observation is not exhaustive. In the example of earthquakes, we know the network reliability. In some sense it can be viewed as a probability of detection, say $\pi(x)$, which is independent of the underlying process of earthquakes. This process is the union of the observed process, $\Phi^\pi$, and the unobserved process, $\Phi^{1-\pi}$.
Then the local intensity is the sum of the intensity of the observed process and the intensity of the unobserved process given the observation, what can be predicted as previously. But now, weights also depend on the observation probability.

The interaction with other processes provides another form of spatial variation. We can easily imagine that the presence or the absence of a species depend on the presence/absence of other species. The extension of our approach to multi-type processes is an on-going work. In that case, we consider several processes, say $\Phi^{(k)}$, $k=1,\dots,K$, observed respectively in some windows $W_1$, ..., $W_K$.
To predict the intensity of the first process given the realization of the others, we can define the predictor as follows
\begin{equation*}
    \widehat{\lambda_1}(x_o |  \Phi^{(1)}_{W_1}, \dots,  \Phi^{(K)}_{W_K}) = \sum_{k=1}^K \sum_{x \in \Phi^{(k)}_{W_k}} w_k(x;x_o).
\end{equation*}
The unbiasedness constraint on the predictor modifies the constraint on the weight function, $$\sum_{k=1}^K \int_{W_k} \lambda_k(x)w_k(x;x_o) \dd x = \lambda_1(x_o),$$
and  minimizing the error prediction variance under this constraint leads to a new Fredholm equation which now depends on the pair correlation function of each process and on the cross-pair correlation functions.

A question remains: how to control the approximation accuracy when solving the Fredholm equation?
There are two ways to increase the accuracy of the computed solution of the Fredholm equation: either refine the mesh, i.e. lower the size of the triangles, or use another approximation basis than the usual one (P1=$\{$\text{piece-wise Linear functions}$\}$).
Both options significantly increase  the dimension of the resulting dense linear problem and from the point of view of computing efficiency we can wonder which is the most suitable approximation. Note that for a given mesh,
we get a more precise  representation of the regularity of the solution if we use a finite element basis involving more regular functions e.g. P2=$\{$\text{piece-wise Quadratic functions}$\}$ or P3=$\{$\text{piece-wise Cubic functions}$\}$\dots However, changing the basis of approximation won't reduce the approximation errors made with respect to the true solution of the Fredholm equation, errors that are intrinsically linked to the typical size of a triangle of the mesh.
Up to our knowledge, no rule has been published that relates the size of the mesh to the characteristics of the equation. A too crude size can even lead to negative estimated local intensities.
We propose to verify empirically the approximation quality by simulating  point realizations using an easy to manipulate point process with the same first- and second-order characteristics, as the log-Gaussian Cox point process, and check discrepancy between reestimated local intensities and the estimated one.



\section*{Acknowledgements}

We thank Giada Adelfio (Universita degli Studi di Palermo), and Antonino D'Alessandro (Istituto Nazionale di Geofisica e Vulcanologia, National Earthquake Center, Rome) for sharing the data with us.

%
\section*{Conflict of interest}
 The authors declare that they have no conflict of interest.

\bibliographystyle{spbasic}      
\bibliography{RefpaperGRCM}

\resub{
\appendix

\section{Proof of Equation~(\ref{Fredhomeq})}
\label{app:fred}

As in \cite{gabriel2017}, the local intensity at locations $x_o \in W \backslash W_{obs}$ given $\Phi\cap W_{obs} = \Phi_{W_{obs}}$  is defined by the limit
\begin{equation*}
\lambda(x_o|\Phi\cap W_{obs} = \Phi_{W_{obs}})=\lim\limits_{\nu(\dd x_o) \to 0}\frac{\bE[\Phi(\dd x_o)|\Phi\cap W_{obs} = \Phi_{W_{obs}}] }{\nu(\dd x_o)},
\end{equation*}
where $\dd x_o$ is an elementary surface around $x_o$.
The spatial predictor of the local intensity is given by
\begin{equation*}
\hat{\lambda}(x_o|\Phi\cap W_{obs} = \Phi_{W_{obs}})=\sum_{x\in \Phi_{W_{obs}}} w(x; x_o).
\end{equation*}
The proof of Equation~(\ref{Fredhomeq})
is similar to that in the stationary setting. For sake of convenience, in what follows, $w(x)$ stands for $w(x;x_o)$ and conditioning over $\Phi\cap W_{obs} = \Phi_{W_{obs}}$ is written $ |\Phi_{W_{obs}}$.

\medskip
The constraint over the spatial weights in order to obtain an unbiased prediction can be expressed as
\begin{equation*}
\bE\left[\hat{\lambda}(x_o|\Phi_{W_{obs}})-\lambda(x_o|\Phi_{W_{obs}})\right]=0,
\end{equation*}
then under non-stationary assumption it becomes
\begin{equation*}
\int\limits_{W_{obs}}\lambda(x)w(x)\dd  x-\bE\left[\lambda(x_o|\Phi_{W_{obs}})\right]=0,
\end{equation*}
and therefore
\begin{equation*}
\int\limits_{W_{obs}}\lambda(x)w(x)\dd x=\lambda(x_o).
\end{equation*}

On the other hand, the predictor must minimize the error prediction variance,
\begin{multline*}
\var\left[\hat{\lambda}(x_{o}|\Phi_{W_{obs}}) - \lambda(x_{o}|\Phi_{W_{obs}}) \right] = \var\left[\hat{\lambda}(x_{o}|\Phi_{W_{obs}}) \right]
+ \var\left[\lambda(x_{o}|\Phi_{W_{obs}}) \right]  \\
- 2 \cov \left[ \hat{\lambda}(x_{o}|\Phi_{W_{obs}}) , \lambda(x_{o}|\Phi_{W_{obs}}) \right],
\end{multline*}
 with
\begin{equation*}
\var\left[\hat{\lambda}(x_{o}|\Phi_{W_{obs}})\right]=\int\limits_{W_{obs}}\lambda(x)w^2(x)\de x+\int\limits_{W_{obs}\times W_{obs}}\lambda(x)\lambda(y)w(x)w(y)(g(x-y)-1)\de x\de y
\end{equation*}
and
\begin{align*}
\cov&\left[\hat{\lambda}(x_{o}\Phi_{W_{obs}}),\lambda(x_{o}|\Phi_{W_{obs}})\right]=\cov\left[\sum_{x_\in \Phi_{W_{obs}}}w(x),\lim\limits_{\nu(\dd x_o) \to 0}\frac{\bE[\Phi(\dd x_o)|\Phi_{W_{obs}}]}{\nu(\dd x_o)}\right]\\
&=\bE\left[\lim\limits_{\nu(\dd x_o) \to 0}\frac{\bE[\Phi(\dd x_o)|\Phi_{W_{obs}}]}{\nu(\dd x_o)}\sum_{x\in \Phi_{W_{obs}}}w(x)\right]-\lambda(x_{o})\int\limits_{W_{obs}}\lambda(x)w(x)\de x\\
&=\lim\limits_{\nu(\dd x_o) \to 0}\int\limits_{W_{obs}\times W_{obs}}\frac{\bE[\Phi(\dd x_o)|\Phi_{W_{obs}}]}{\nu(\dd x_o)}\lambda(x)\lambda(y)w(x)g(x-y)\de x\de y-\lambda(x_{o})\int\limits_{W_{obs}}\lambda(x)w(x)\de x\\
&=\lambda(x_{o})\int\limits_{W_{obs}}\lambda(x)w(x)(g(x_o-x)-1)\de x.
\end{align*}
This is equivalent to minimise the following expression
\begin{align*}
&\int\limits_{W_{obs}}\lambda(x)w^2(x)\de x+\int\limits_{W_{obs}\times W_{obs}}\lambda(x)\lambda(y)w(x)w(y)(g(x-y)-1)\de x\de y\\
&\qquad\qquad-2\lambda(x_{o})\int\limits_{W_{obs}}\lambda(x)w(x)(g(x_o-x)-1)\de x.
\end{align*}
We consider the Lagrange multipliers under the constraint for the spatial weight function and we define
\begin{align*}
T(w(x))=&\int\limits_{W_{obs}}\lambda(x)w^2(x)\de x+\int\limits_{W_{obs}\times W_{obs}}\lambda(x)\lambda(y)w(x)w(y)(g(x-y)-1)\de x\de y\\
&-2\lambda(x_{o})\int\limits_{W_{obs}}\lambda(x)w(x)(g(x_o-x)-1)\de x+\mu\left(\int\limits_{W_{obs}}\lambda(x)w(x)\de x-\lambda(x_o)\right).
\end{align*}
For $\alpha(x)=w(x)+\varepsilon(x)$, we have
\begin{align*}
T(\alpha(x))=&\int\limits_{W_{obs}}\lambda(x)(w(x)+\varepsilon(x))^{2}\de x\\
&+\int\limits_{W_{obs}\times W_{obs}}\lambda(x)\lambda(y)\left(w(x)+\varepsilon(x)\right)\left(w(y)+\varepsilon(y)\right)(g(x-y)-1)\de x\de y\\
&-2\lambda(x_{o})\int\limits_{W_{obs}}\lambda(x)\left(w(x)+\varepsilon(x)\right)(g(x_o-x)-1)\de x\\
&+\mu\left(\int\limits_{W_{obs}}\lambda(x)\left(w(x)+\varepsilon(x)\right)\de x-\lambda(x_o)\right),
\end{align*}
then
\begin{align*}
T&(\alpha(x))=\int\limits_{W_{obs}}\lambda(x)w^{2}(x)\de x+2\int\limits_{W_{obs}}\lambda(x)w(x)\varepsilon(x)\de x\\
&+\int\limits_{W_{obs}\times W_{obs}}\lambda(x)\lambda(y)w(x)w(y)(g(x-y)-1)\de x\de y\\
&+\int\limits_{W_{obs}\times W_{obs}}\lambda(x)\lambda(y)w(x)\varepsilon(y)(g(x-y)-1)\de x\de y\\
&+\int\limits_{W_{obs}\times W_{obs}}\lambda(x)\lambda(y)w(y)\varepsilon(x)(g(x-y)-1)\de x\de y\\
&-2\lambda(x_{o})\int\limits_{W_{obs}}\lambda(x)w(x)(g(x_o-x)-1)\de x-2\lambda(x_{o})\int\limits_{W_{obs}}\lambda(x)\varepsilon(x)(g(x_o-x)-1)\de x\\
&+\mu\left(\int\limits_{W_{obs}}\lambda(x)w(x)\de x+\int\limits_{W_{obs}}\lambda(x)\varepsilon(x)\de x-\lambda(x_o)\right)+2O(\varepsilon(x)).
\end{align*}
Finally, we can rewrite the previous expression as
\begin{align*}
T&(\alpha(x))\approx T(w(x))+2\int\limits_{W_{obs}}\lambda(x)w(x)\varepsilon(x)\de x+2\int\limits_{W_{obs}\times W_{obs}}\lambda(x)\lambda(y)w(y)\varepsilon(x)(g(x-y)-1)\de x\de y\\
&-2\lambda(x_{o})\int\limits_{W_{obs}}\lambda(x)\varepsilon(x)(g(x_o-x)-1)\de x+\mu\int\limits_{W_{obs}}\lambda(x)\varepsilon(x)\de x+2O(\varepsilon(x)),
\end{align*}
and therefore
\begin{align*}
&T(\alpha(x))\approx T(w(x))+2\int\limits_{W_{obs}}\varepsilon(x)\Bigg[\lambda(x)w(x)+\int\limits_{W_{obs}}\lambda(x)\lambda(y)w(y)(g(x-y)-1)\de y\\\
&\qquad\qquad\qquad\qquad\qquad-\lambda(x_{o})\lambda(x)(g(x_o-x)-1)+\frac{\mu}{2}\lambda(x)\Bigg]\de x,
\end{align*}
Using variational calculation and the Riesz representation theorem, it follows that
\begin{equation*}
T(\alpha(x))-T(w(x))= O(\varepsilon(x)),
\end{equation*}
that leads to
\begin{align*}
&2\int\limits_{W_{obs}}\varepsilon(x)\Bigg[\lambda(x)w(x)+\int\limits_{W_{obs}}\lambda(x)\lambda(y)w(y)(g(x-y)-1)\de y\\
&\qquad\qquad\qquad\qquad\qquad\qquad\qquad-\lambda(x_{o})\lambda(x)(g(x_o-x)-1)+\frac{\mu}{2}\lambda(x)\Bigg]\de x=0,
\end{align*}
and therefore
\begin{align}
\label{eq:fred0}
&\lambda(x)w(x)+\int\limits_{W_{obs}}\lambda(x)\lambda(y)w(y)(g(x-y)-1)\de y\\ \nonumber
&\qquad\qquad\qquad\qquad\qquad\qquad\qquad-\lambda(x_{o})\lambda(x)(g(x_o-x)-1)+\frac{\mu}{2}\lambda(x)=0.
\end{align}
Considering the integral of previous equation over the spatial window $W_{obs}$ and respect to $x$, it follows that
\begin{align*}
&\int\limits_{W_{obs}}\lambda(x)w(x)\de x+\int\limits_{W_{obs}\times W_{obs}}\lambda(x)\lambda(y)w(y)(g(x-y)-1)\de x\de y\\
&\qquad\qquad\qquad\qquad\qquad\qquad-\lambda(x_{o})\int\limits_{W_{obs}}\lambda(x)(g(x_o-x)-1)\de x+\frac{\mu}{2}\int\limits_{W_{obs}}\lambda(x)\de x=0,
\end{align*}
and then we obtain
\begin{align*}
&\lambda(x_o)+\int\limits_{W_{obs}\times W_{obs}}\lambda(x)\lambda(y)w(y)(g(x-y)-1)\de x\de y\\
&\qquad\qquad\qquad\qquad\qquad\qquad-\lambda(x_{o})\int\limits_{W_{obs}}\lambda(x)(g(x_o-x)-1)\de x+\frac{\mu}{2}\int\limits_{W_{obs}}\lambda(x)\de x=0,
\end{align*}
from which we get
\begin{align*}
\frac{\mu}{2}=&\frac{1}{\int\limits_{W_{obs}}\lambda(x)\de x}\Bigg[\lambda(x_{o})\int\limits_{W_{obs}}\lambda(x)(g(x_o-x)-1)\de x\\
&\qquad\qquad\qquad\qquad\qquad-\int\limits_{W_{obs}\times W_{obs}}\lambda(x)\lambda(y)w(y)(g(x-y)-1)\de x\de y-\lambda(x_o)\Bigg].
\end{align*}
Finally, plugging $\mu/2$ into Equation~(\ref{eq:fred0}), it follows that
\begin{align*}
&\lambda(x)w(x)+\int\limits_{W_{obs}}\lambda(x)\lambda(y)w(y)(g(x-y)-1)\de y\\
&\qquad-\lambda(x_{o})\lambda(x)(g(x_o-x)-1)+\frac{\lambda(x)}{\int\limits_{W_{obs}}\lambda(z)\de z}\Bigg[\lambda(x_{o})\int\limits_{W_{obs}}\lambda(z)(g(x_o-z)-1)\de z\\
&\qquad\qquad\qquad\qquad-\int\limits_{W_{obs}\times W_{obs}}\lambda(z)\lambda(y)w(y)(g(z-y)-1)\de z\de y-\lambda(x_o)\Bigg]=0,
\end{align*}
and after grouping we arrive to the Fredhom equation of second kind
\begin{align*}
&\lambda(x)w(x)+\lambda(x)\int\limits_{W_{obs}}\lambda(y)w(y)(g(x-y)-1)\de y\\
&\qquad-\frac{\lambda(x)}{\int\limits_{W_{obs}}\lambda(z)\de z}\left[\lambda(x_o)+\int\limits_{W_{obs}\times W_{obs}}\lambda(z)\lambda(y)w(y)(g(z-y)-1)\de z\de y\right]\\
&\qquad\qquad=\lambda(x_{o})\lambda(x)(g(x_o-x)-1)-\frac{\lambda(x_{o})\lambda(x)}{\int\limits_{W_{obs}}\lambda(z)\de z}\int\limits_{W_{obs}}\lambda(z)(g(x_o-z)-1)\de z
\end{align*}

\section{Zoom of predictions}
\label{app:zoom}

Figures~\ref{fig:zoompred} and \ref{fig:zoompredH} correspond to the middle and right panels of Figures~\ref{fig:patterns} and \ref{fig:patternsH} zoomed on the prediction window.

\begin{figure}[h]
  \hspace{4cm}$\lambda(x_o | \Phi_{w_{obs}})$ \hspace{1.5cm} $\widehat \lambda(x_o | \Phi_{w_{obs}})$

      \vspace{2mm}

    \centering
  \epsfig{file=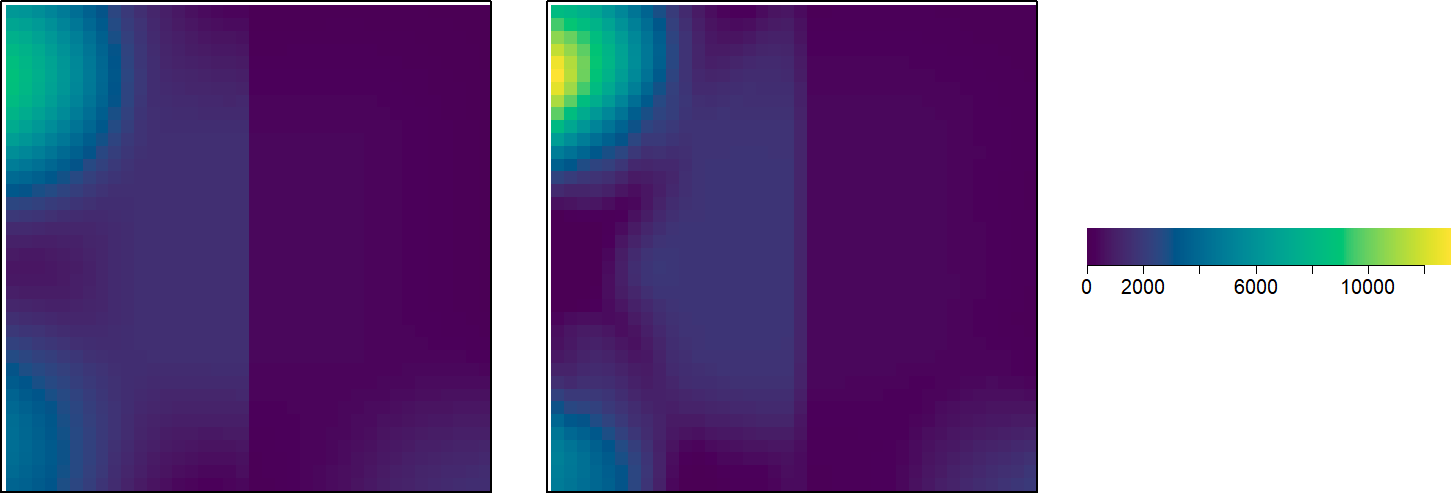,width=0.5\textwidth}

      \vspace{2mm}
    \epsfig{file=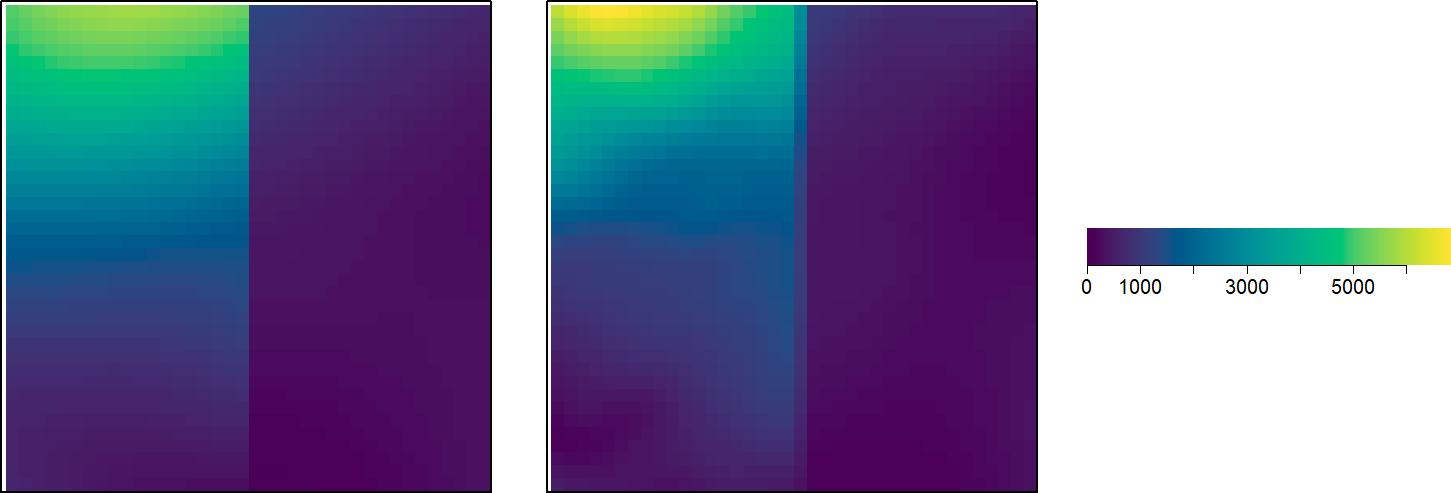,width=0.5\textwidth}

      \vspace{2mm}
      \epsfig{file=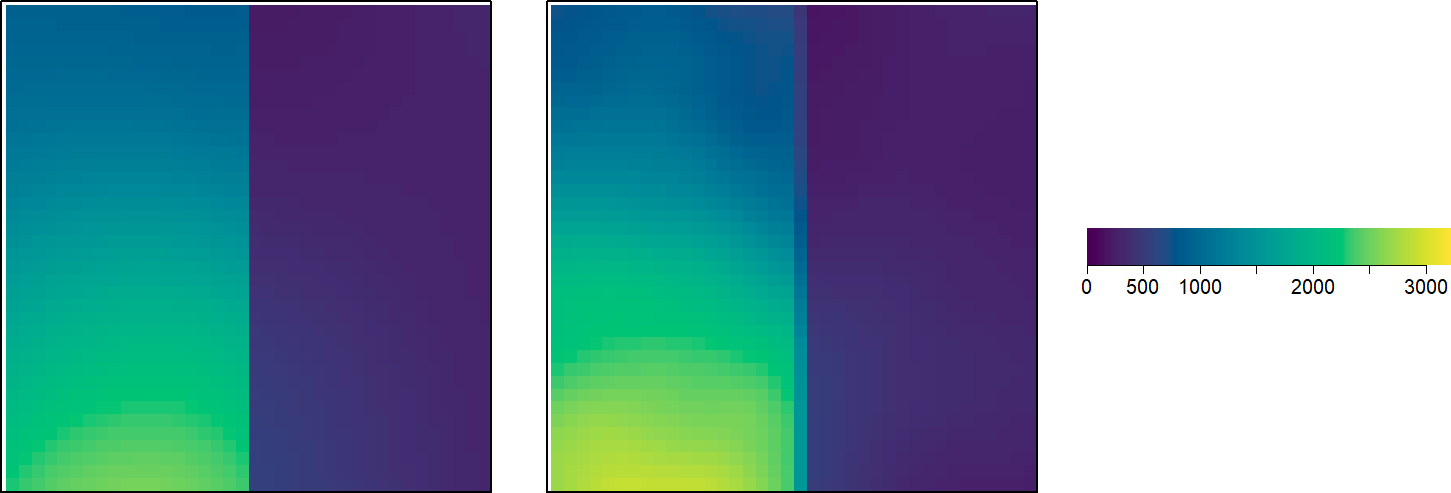,width=0.5\textwidth}
  \caption{Conditional intensity (left panels) and prediction of the local intensity (right panels)  obtained from IMCP$(p_1(x),R)$ with $R=0.05$ (first row), $R=0.09$ (second row) and $R=0.013$ (third row).}\label{fig:zoompred}
\end{figure}

\begin{figure}[h]
  \hspace{3.25cm}$\lambda(x_o | \Phi_{w_{obs}})$ \hspace{3.5cm} $\widehat \lambda(x_o | \Phi_{w_{obs}})$

      \vspace{2mm}
    \centering
  \epsfig{file=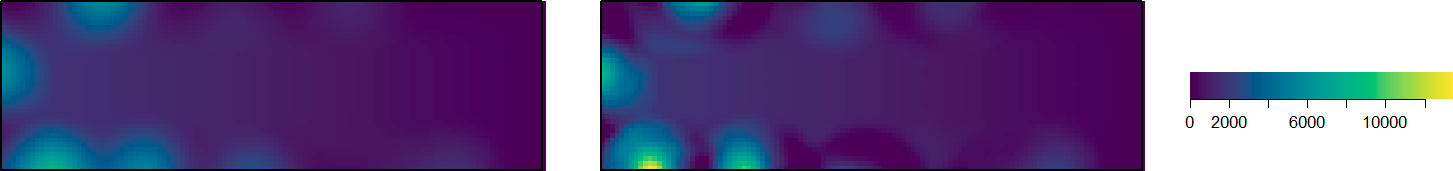,width=0.65\textwidth}

  \vspace{2mm}
    \epsfig{file=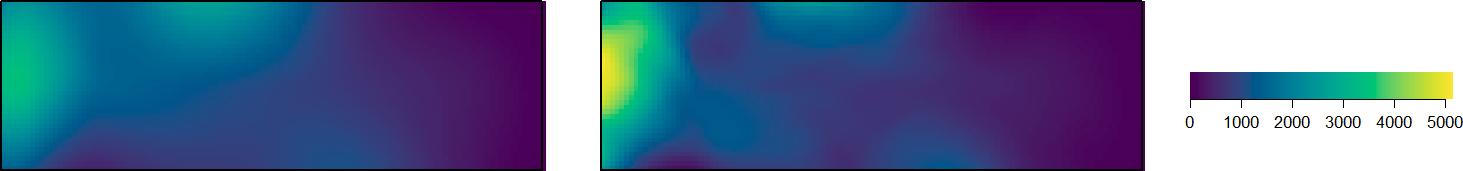,width=0.65\textwidth}

      \vspace{2mm}
      \epsfig{file=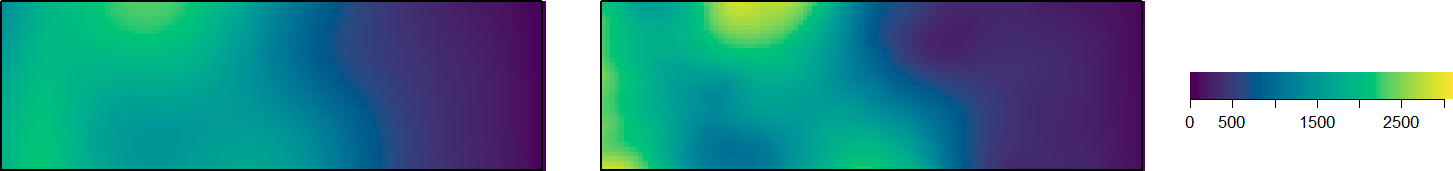,width=0.65\textwidth}
  \caption{Conditional intensity (left panels) and prediction of the local intensity (right panels)  obtained from IMCP$(p_2(x),R)$ with $R=0.05$ (first row), $R=0.09$ (second row) and $R=0.013$ (third row).}\label{fig:zoompredH}
\end{figure}

\section{First- and second-order moments estimation}
\label{app:estimation}

In section~\ref{sec:simulation} we analyze the effect of plugging estimates of  the intensity and the pair correlation functions in the predictor.
We considered a parametric model for the intensity function, $f_\lambda (x)$, two parametric (Mat{\'e}rn and Exponential) and one non parametric (referred to as empirical and denoted $g_{emp}(r)$) models for the pair correlation function.
 Table~\ref{table:param} summarizes the fitted parameters of the different models.
\begin{table}[h]
\centering
\caption{Estimated parameters of the intensity model $f_\lambda$ and of the Mat{\'e}rn and exponential models for the pair correlation function. Mean values are in bold, standard deviation in italic. }
\scalebox{1.0}{
\begin{tabular}{|cc|cc|cc|}
  \hline
  \multicolumn{2}{|c|}{$f_\lambda(x)$} & \multicolumn{2}{|c|}{$g_{mat}(r)$} & \multicolumn{2}{|c|}{$g_{exp}(r)$} \\
  \hline
   $\beta_1$ & $\beta_2$ & $\alpha_1$ & $\alpha_2$ & $\alpha_3$ & $\alpha_4$ \\
  \hline
  {\small \textbf{408} (\textit{59})} & {\small  \textbf{1648} (\textit{245})} & {\small \textbf{53.7} (\textit{15.3})} & {\small \textbf{0.078} (\textit{0.014})} & {\small \textbf{0.377} (\textit{0.073})} & {\small \textbf{8.69} (\textit{1.74})} \\
  \epsfig{file=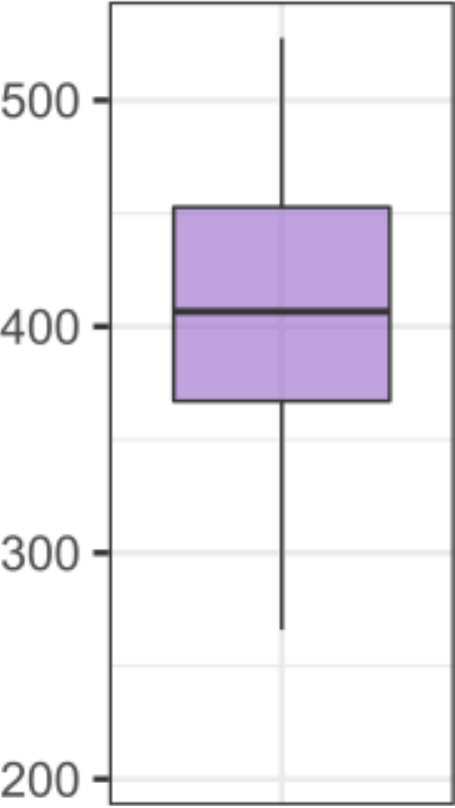,width=1.7cm} & \epsfig{file=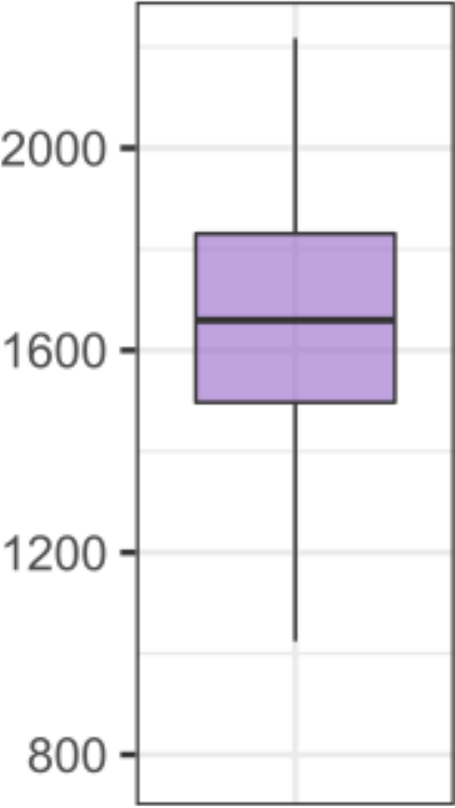,width=1.7cm} &  \epsfig{file=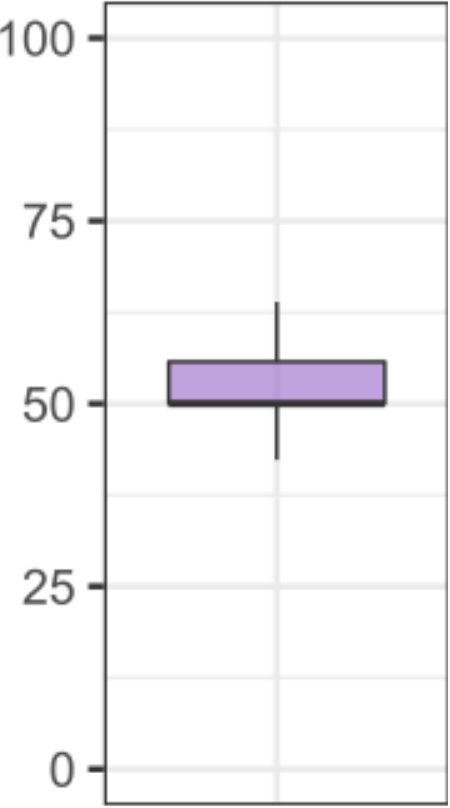,width=1.7cm} & \epsfig{file=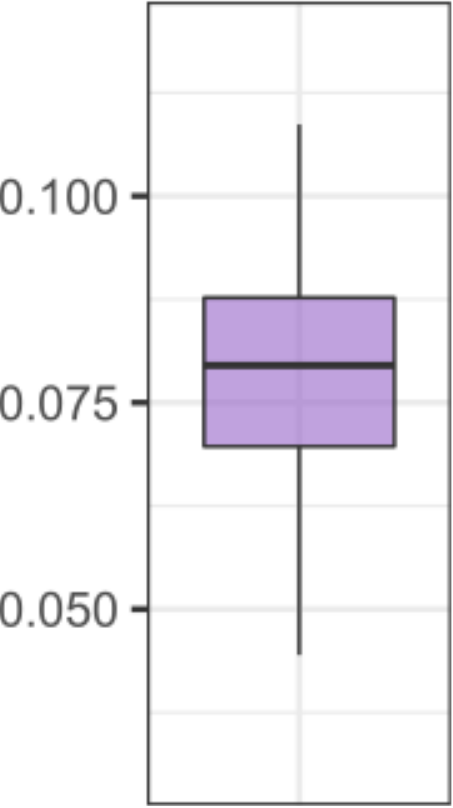,width=1.7cm}&  \epsfig{file=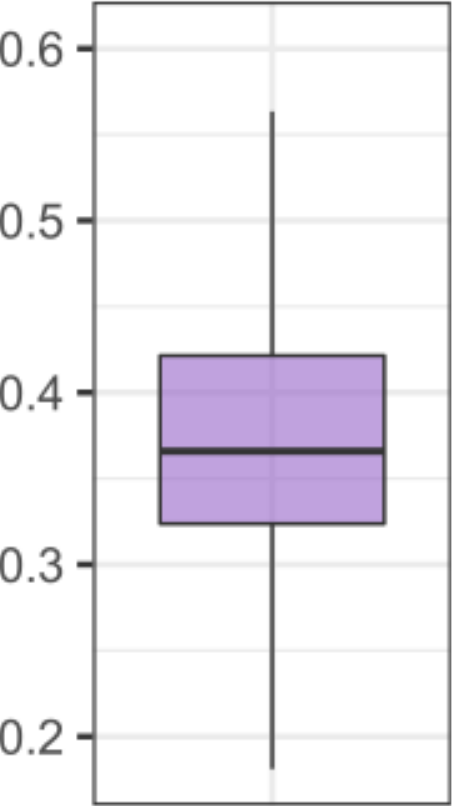,width=1.7cm} & \epsfig{file=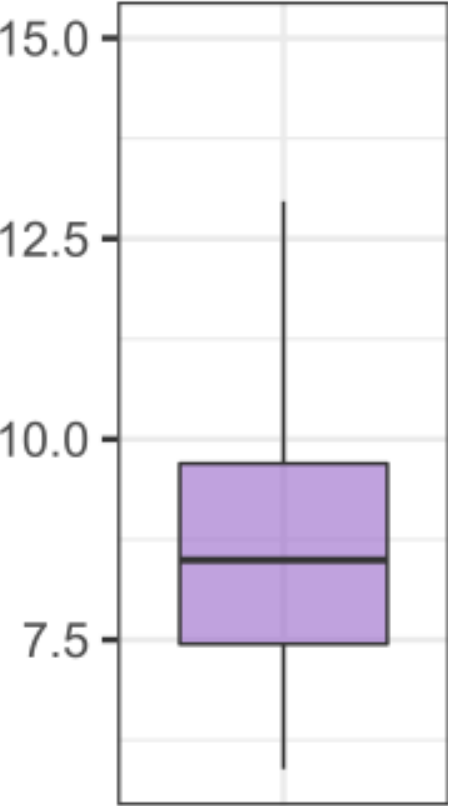,width=1.7cm} \\
  \hline
\end{tabular}}
\label{table:param}
\end{table}

The non parametric estimate of the pair correlation function is obtained by kernel smoothing, as implemented in the \texttt{pcfinhom} function in \texttt{spatstat} with $f_\lambda (x)$ as intensity function.
However this function may lead to odd estimates that often do not satisfy theoretical conditions of a pair correlation function (positive definiteness, consistence). Odd estimates are plotted in Figure~\ref{fig:badpcf} and show that they do not tend to 1 as $r$ increases.
\begin{figure}
  \centering
\epsfig{file=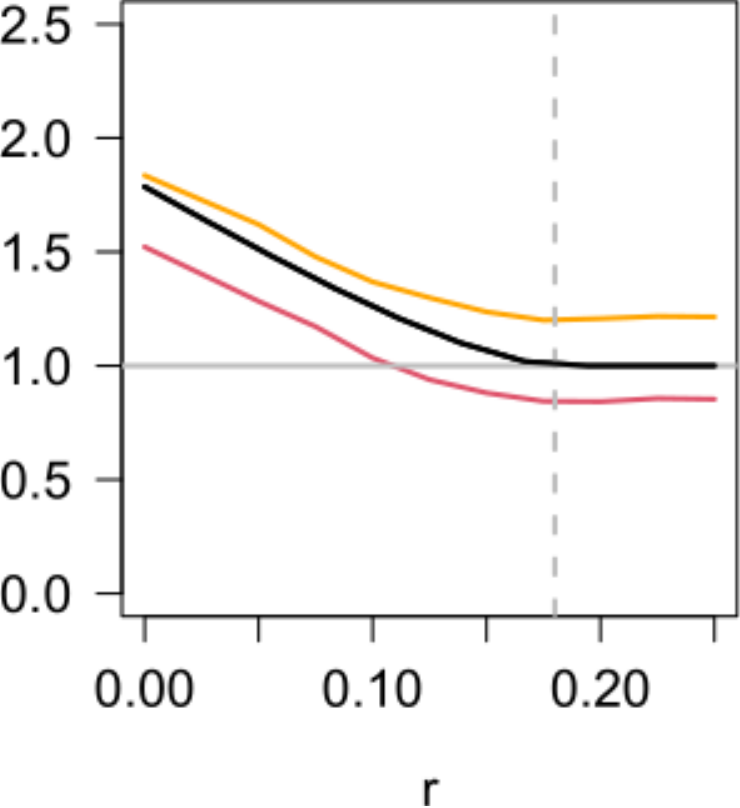,width=0.35\textwidth}
  \caption{Examples (orange and red curves) of odd non parametric estimates of the pair correlation function. The theoretical model is in black. The vertical dashed line corresponds to the distance $2R$.}\label{fig:badpcf}
\end{figure}
 Regarding the expression of the weights (Equation~\ref{Fredhomeq}), this highly impact the predictions: we get very hight values for the related simulations. For instance from the estimates plotted in  Figure~\ref{fig:badpcf} we get predictions between $-10130$ and $13295$.
Hence for the simulation study, we made an \textit{a posteriori} selection of the admissible fitted pair correlation functions as follows~:
\begin{itemize}
  \item $\text{median}\lce | g_{emp}(r) - 1 |; \! r>2R \rce < 0.05$, i.e.
  the median distance between $g_{emp}(r)$ and 1 for all $r > 2R$ must be less than 0.05,
  \item the predictions must be positive and not exceed 6000.
\end{itemize}

}

\end{document}